\documentclass[prb,twocolumn,showpacs]{revtex4}
\usepackage{times}

\newcommand{\be}{\begin{equation}}
\newcommand{\ee}{\end{equation}}
\newcommand{\bea}{\begin{eqnarray}}
\newcommand{\eea}{\end{eqnarray}}
\newcommand{\rd}{\mbox{d}}
\newcommand{\p}{\partial}
\newcommand{\textbinm}[2]{{\textstyle {#1\choose #2}}}

\usepackage{graphicx}
\usepackage{dcolumn}
\usepackage{bm}

\begin{document}

\title{      
Interedge coherent line junctions in Quantum Hall systems}

\author{Emiliano Papa} 

\address{Department of Physics, The University of Virginia, Charlottesville, VA 22904}
\date{\today}

\begin{abstract}
In this paper we address some of the properties of quantum Hall line junctions (QHLJ) that occur near barriers separating electron gases on 
quantum Hall plateaus.
In narrow barriers where electron tunneling can occur, the low energy physics of the QHLJ is described by the quantum sine-Gordon model.
We propose procedures to study a sort of properties of these systems in relation with recent experimental studies of Kang {\it et al.}, 
Nature {\bf 403}, 59 (2000).  
We propose experimental ways of measuring the values of the dimensionless coupling constant characteristic for the Sine-Gordon 
model as well as its Fermi velocity.
When the bulk filling of the 2DEG subsystems is $\nu \sim 2$, these systems can be good candidates for observation of spin-incoherent LL behavior.

\end{abstract}


\maketitle

\section{Introduction}

The edge states of quantum Hall systems offer a rich ground 
for testing the theoretical predictions on the properties in one spatial dimension (1D) of strongly interacting fermionic systems. 
Unlike in quantum wires on which impurities can lead to localization effects hindering the 
observation of the generic properties of 1D fermions, 
the QH edge states are considered to be very clean realizations of the Luttinger liquid description. 
The QHLJ that occur near disorder free barriers separating electron systems on quantum Hall (QH) plateaus, have been shown to be 
an important tool on these studies.

Properties of QHLJ, as the differential tunneling conductance between counter-propagating edge states 
has been studied extensively in experiment by Kang and coworkers \cite{kang,kang04-05}, by Grayson {\it et al.,} 
\cite{graysonapl,graysonprl}, and by Roddaro {\em et al.,}\cite{pellegrini} as well as in theory \cite{renn,mitra,kollar,kim}. 
In the experiments  of Refs.~\onlinecite{kang,kang04-05} the two-dimensional electron gases (2DEG) are separated by a narrow barrier, 
constructed with the
atomically precise cleaved-edge overgrowth (CEO) technique. Such narrow barriers enable strong correlations between 
electrons on opposite sides of the junction. 
At finite barrier heights the intersections of the Landau levels
in the barrier region turn into gaps. 
When $\nu$ falls within the gap, transport along the barrier is forbidden and the QH effect extends throughout the sample. 
In Ref.~\onlinecite{kang} the tunneling conductance across the barrier displays zero bias peaks of amplitude $G\sim 0.1 e^2/h$ 
over a range of filling fractions $\delta \nu \approx 0.3$ centered at $\nu^*=1.35$ [as well as at points between other higher integer fillings].
The size of the gap ($\Delta \sim \delta \nu\, \hbar \omega_c$)  however has been a point for which the theoretical 
\cite{mitra,kollar} and experimental \cite{kang} results have been in partial disagreement. 

In Ref.~\onlinecite{mitra} the formation  of a phase-coherent state of fermions between opposite sides of the barrier has 
been proposed for the explanation of the $I$-$V$ characteristics of this system.
According to this model, in the presence of fermionic tunneling between edges on opposite sides of the barrier, the low energy 
physics of the 
QHLJ is expected to be described by the quantum sine-Gordon (SG) model in terms of variable $\varphi$ that represents the phase of the 
interedge phase-coherence order parameter.   
The properties of the SG model are dependent on the value of the coupling constant $\beta$ and   
the value of the Fermi velocity. 
The accuracy of estimation of the physical mass in this quantum field theory however suffers from two limitations. 
First, the SG parameters $\beta$ and $v_F$ appearing in the bosonized version that describes the line junction at low energies 
are hard to be estimated with accuracy from microscopic models. Second, the physical mass 
depends also on the choice of the ultraviolet cutoff regulating the 
short distance behavior of the quantum field theory. Progress in the latter direction has been achieved by using the method 
of asymptotic matching \cite{kollar}, in 
which the long distance of the quantum field theory and the short distance of the microscopic model are 
matched in the region of common validity. 
The theoretical calculations for the magnitude of the gap within this model give an estimate of
$\Delta \sim 1.3 {\rm K}$ (Ref.~\onlinecite{mitra}) and 
$\Delta \sim 1.5{\rm K}$ (Ref.~\onlinecite{kollar}). In the latter calculation, the increased value of magnitude of the gap 
was attributed to the renormalization of the tunneling amplitude by interactions as well as to the better estimation of the 
short-distance cutoff of the model.
The experimental value of the energy gap above the ground state measured in the experiments of Ref.~\onlinecite{kang} 
is one order of magnitude higher  $\Delta \sim 15 {\rm K}$.
To explain the features of the experiment, in Ref.~\onlinecite{kim}
the model with single strong tunneling center along the barrier was proposed.
Wide barriers due to differences in the Fermi wave-vectors of the states on opposite sides of the barrier, 
do not allow for tunneling to take place at any point except at imperfection ones 
created unavoidably during the fabrication process. 
Narrow barriers on the other hand can allow for such tunneling to take place 
at any point along the barrier. 
Our work here is intended for the case when tunneling can take place at an infinite number of points along the barrier. 
The results and predictions of this work if examined in the experiments can however shed light on the model that best describes the experiments.  

Assuming therefore the model where tunneling can take place at an infinite number of points along 
the barrier here we show that one can indirectly extract information on the value of these 
quantities ($\beta$ and $v_F$) by using the integrability property of the SG model. 
The spectrum of the quantum sine-Gordon model consists of massive topological particles, solitons, 
antisolitons, and when forward interactions are strong enough also of their bound states. In 
the presence of a chemical potential however that couples 
with these charges the spectral gap can be suppressed. This phenomenon is similar to the one that is 
discussed previously in the context of quantum ferromagnetism and phase transitions in bilayer QH systems 
in the presence of a magnetic field coplanar with the 2DEG layers  and also interlayer 
tunneling experiments in bilayer QH systems in the presence of 
a magnetic field coplanar with the 2DEG layers (see Refs.~\onlinecite{yang-94,girvin-2001}). When this 
chemical potential exceeds the mass of the soliton ($\Delta/2$), a finite density of solitons appears in the ground state,
distributed on a Fermi sea according to their statistics, embedded in their interactions (or their scattering matrix).
The low-energy physics of this system then will be of particle-hole type formed around these Fermi points. The properties of 
this metallic state, namely the value of the Luttinger liquid (LL) parameter $K$ and the Fermi velocity can be accessed with the 
thermodynamic Bethe ansatz. 
It turns out that at values of chemical potential not very far from $\Delta$, the LL parameters 
$K$ and $v_F$ approach the values of the 
SG coupling constant $\beta$ and $v_{F}$ prior to gap opening.
Experimentally there are two quantities that offer the measurement of two combinations, the 
product and the ratio of the LL parameter $K$ 
and the Fermi velocity, namely the Drude weight and charge susceptibility, respectively. 

On the other hand, in the insulating phase of the model we propose the measurement 
of the optical conductivity of the system. 
The optical conductivity measurements should measure a delta 
function-like presence for each of the kinds of particles of the spectrum odd under charge conjugation
and also a
contribution on the form of a continuous band for their particle-particle combinations 
(with specific selection rules). 
The proposed measurement of the optical conductivity therefore would be an important confirmation of 
the existence of the interedge coherent state.
It would serve also the purpose of bounding the value of the coupling constant.

The SG model has applications in other contexts too where therefore it would be helpful to know when the 
finite temperature LL description starts to become unreliable. 
This is something that we also discuss in this paper.

At filling fractions $\nu \sim 2$ the spin degrees of freedom play a role in the dynamics of the system. In this case terms like spin exchange 
interactions as well as the coupling of the spin current to the magnetic field lead to a rich physics. As discussed in Ref.~\onlinecite{kim},
the spin sector can be in a gapped state. This gap can be suppressed in the presence of the magnetic field. 
At values of this field only slightly in excess of the gap the energy scales in the spin and the 
charge sectors separate. At a small window of energies above the spin gap at finite temperatures 
the spin incoherent LL (SILL) state \cite{fiete} can be realized. Experimental implications on charge 
and spin transport are discussed.

\section{The model} 

\begin{figure}
\unitlength=1mm
\begin{picture}(80,41)
\put(-3,3.2){ \includegraphics[width=34.33\unitlength]{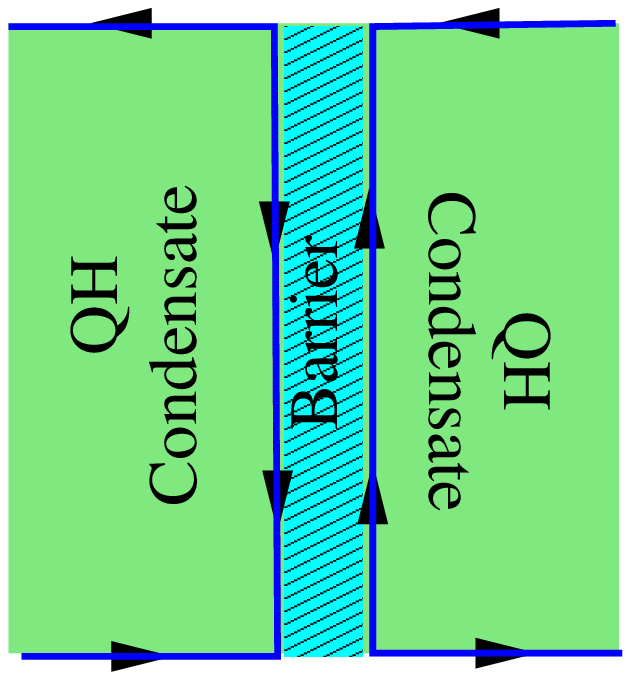}}
\put(35.,-4){\includegraphics[width=48\unitlength]{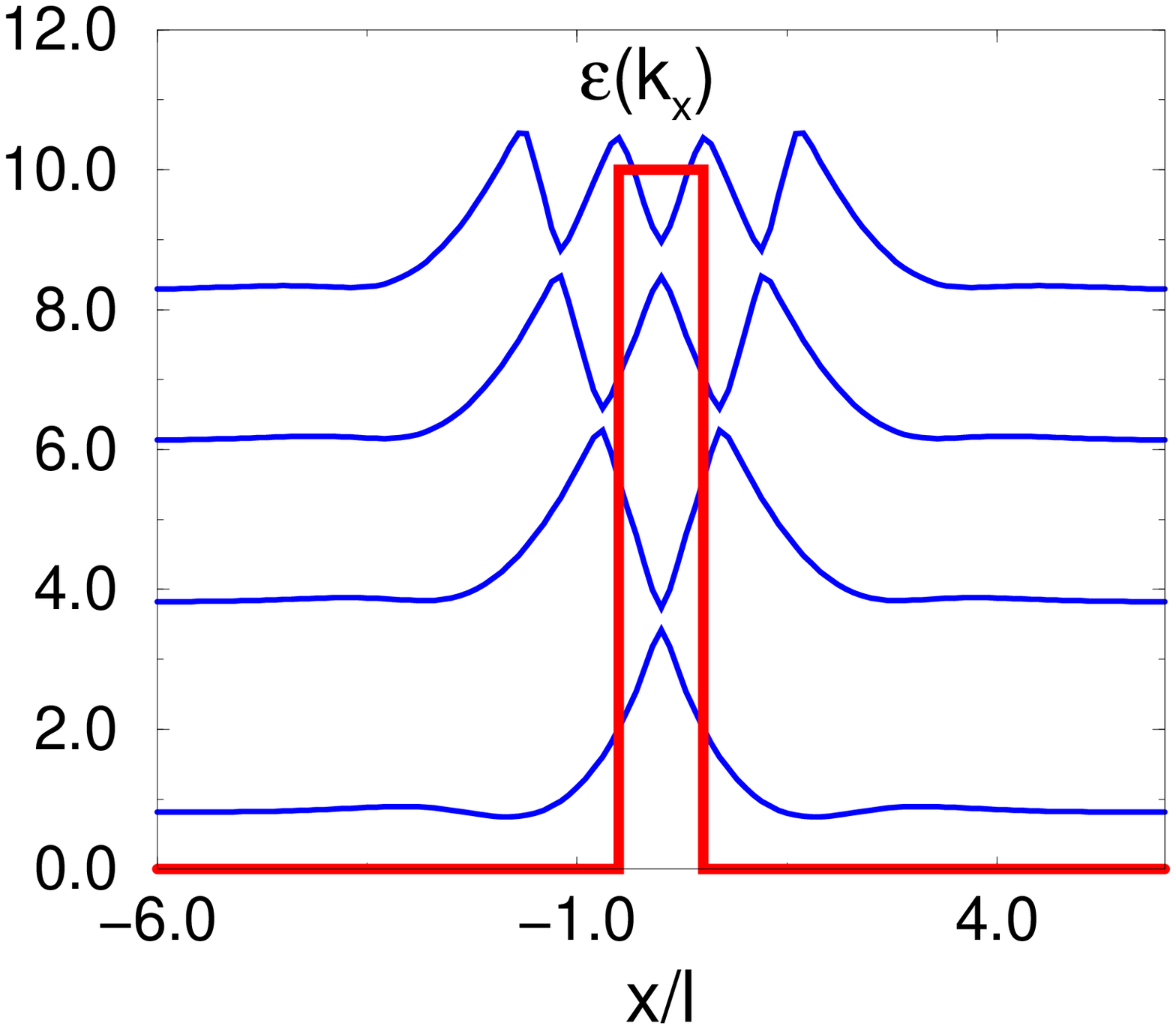}} 
\end{picture}
\caption{ Left: Schematic illustration of a QHLJ formed in QH systems by means of cleaved edge 
overgrowth barriers, similar to the ones used in experimental studies of Kang and coworkers 
\protect\cite{kang}. The barrier is considered to be narrow enough to allow for interedge tunneling 
to take place at an infinite number of points. Right: the electron dispersion curves $\varepsilon(k_x)$ 
in the presence of interactions are shown. In mean field Hartee-Fock theory the dispersion relation 
of electrons can be found by solving the equation
$
\left\{-\frac{\hbar^2}{2m}\p_y^2  +  
\frac{m \omega_c^2}{2} (y-l^2 k)^2 +V^H(y) + V_B(y)\right\}\chi_k(y)
-\int {\rd } {y'}\, V^E(y,y')\chi_k(y') = \varepsilon_k\chi_k(y),
$
where $V^E(y,y') = (2/L_x)\sum_k' \chi_{k'}^\star(y')\chi_{k'}(y){\bf K}(k-k')(y-y'))$ is the exchange interaction experienced 
by the electrons on the same side of the barrier, $V^H(y)=(-1/L_x)\sum_{k'}\int {\rd}{y'}\,  |\chi_{k'}(y')|^2 \ln[((y-y')^2+w^2)/L_x^2]$, 
is the direct interaction, and $V_B(y)$ is the potential presented to the electrons by the barrier.  $w$ is a short distance cutoff 
of the order of the magnetic length.  At finite barrier height, energy gaps appear in the barrier region.}
\label{fig:QHLJ-model}
\end{figure}

In this section we introduce the model that intends to describe 
the experimental systems used to study properties of QHLJ. 
The geometry of the model is represented schematically in  Fig.~\ref{fig:QHLJ-model}.
A QH system at a plateau at integer filling fraction $\nu=1$ is separated down the middle by a narrow barrier as illustrated  
in Fig.~\ref{fig:QHLJ-model}. The barrier creates on 
its opposite sides edge states that propagate in opposite directions. The potential that CEO barriers present to the electron gas in the experimental systems
are sufficiently abrupt that
edge reconstruction phenomena and complications in their description do not arise.
Therefore we describe the edges by single chiral moving channels.
At the integer filling fraction $\nu=1$ the spin degree of freedom is frozen 
and does not contribute in the dynamics of the system at low energies.
 
The QH samples used in Ref.~\onlinecite{kang} are long in the direction of the barrier. More specifically,
the lateral dimensions of the QH samples used 
are $\sim 100 \mu m$ long in the direction of the line junction (noted $L_x$ below) and 13--14$\mu m$ wide from each side 
of the junction.  The 2D electron systems are separated from each-other by a 8.8nm thick barrier. 

The electron states on each side of the barrier 
can be described by single-particle wave functions extended in the direction
of the barrier $\psi_k({\bf r}) = e^{ik_x x}\chi_k({\bf r})/\sqrt{L_x}$, chosen here to be the 
$\hat{x}$ direction, and labeled by an one-dimensional wave vector $k_x$
that is proportional to the guiding center along which the wave function's $y$-coordinate is localized, $Y=k_xl^2$.
In mean field Hartee-Fock theory the dispersion relation of electrons can be found by solving the equation
$
\left\{-\frac{\hbar^2}{2m}\p_y^2 + 
\frac{m \omega_c^2}{2} (y-l^2 k)^2 +V^H(y) + V_B(y)\right\}\chi_k(y)
-\int {\rd} {y'}\, V^E(y,y')\chi_k(y') = \varepsilon_k\chi_k(y), 
$
where $V^E(y,y')$ is the exchange interaction experienced by the electrons on the same side of the barrier 
(described by the Slater determinant). 
where $V^E(y,y') = (2/L_x)\sum_k' \chi_{k'}^\star(y')\chi_{k'}(y){\bf K}(k-k)'(y-y'))$ is the exchange interaction experienced
by the electrons on the same side of the barrier and 
$V^H(y)=(-1/L_x)\sum_{k'}\int {\rd}y'\, |\chi_{k'}(y')|^2 \ln[((y-y')^2+w^2)/L_x^2]$ is the direct interaction. $V_B(y)$ is the 
potential presented to the electrons by the barrier, whereas
$w$ is a short distance cutoff of the order of the magnetic length.
The dispersion relations $\varepsilon(k_x)$ and the corresponding gaps that appear in the intermixing region, are shown in Fig.~\ref{fig:QHLJ-model}, right.

The classical quadratic fluctuations of the coupled chiral edges on the sides of the barrier
are described by a Hamiltonian density of the form $H\sim u_\alpha (x) K_{\alpha\beta}(x-x') \, u_\beta(x')$, 
where summation for repeated indices is assumed. $u_\pm$ are the transverse displacements of the right 
and left edge, respectively, from the ground state location. 
In principle these interaction kernels can be calculated by using of the Hartree-Fock mean field theory.

Adopting field theory terminology the quadratic Hamiltonian can be written in terms of chiral currents
$J_\alpha(x)=  u_\alpha(x)/(2\pi l^2)$,
where $\alpha=\pm1$, for the current on the right and the left of the junction, respectively
\begin{equation}
\label{Hamiltonian_2nw}
H = \frac{\pi v_F}{2} \int \rd x \; \rd x' \sum_{\alpha,\beta=\pm}J_\alpha(x) K_{\alpha\beta}J_\beta(x')
\;.
\end{equation}
The Hamiltonian can be rewritten in terms of sums and differences of currents of the edges on the left and on the right of the barrier, 
with the corresponding kernels also given by sums and differences of the
kernels of same and different sides, 
$K_\gamma(x-x')= l^2[K_{RR}(x-x')+\gamma K_{RL}(x-x')]$, where $\gamma=\pm$.
Since the barrier's thickness is of the order of the magnetic length, the Coulomb interaction between 
fermions on the same side of the barrier $K_{RR}$ and $K_{LL}$, where $K_{RR} = K_{LL}$, 
are comparable to the one between fermions on opposite sides of the barrier $K_{RL}$.
Therefore one of the kernels 
$K_-$ ($K_\Theta$ below) should be expected to be much softer than the other.
The Hamiltonian (\ref{Hamiltonian_2nw}) is quantized by requiring for the chiral currents to 
fulfill the commutation relations
\bea
\left[J_\alpha(x),J_\beta(x')\right]= \alpha \frac{i}{2\pi}\delta_{\alpha,\beta} \p_x \delta(x-x')
\quad .
\eea

The chiral currents $J(x)=:R^\dagger(x)R(x):$ and $\bar J(x)=:L^\dagger(x)L(x):$, 
where $R^\dagger$ and $L^\dagger$ are the fermion creation operators on the right and left moving edge, respectively, can be expressed in terms of bosonic fields
\bea
J(x) =-\frac{1}{\sqrt{\pi}} \p_x \phi_R(x)\quad, \quad \bar J(x)=-\frac{1}{\sqrt{\pi}} \p_x \phi_L(x)\quad.
\eea
Here we used the convention $R(x)\sim e^{-i \sqrt{4\pi} \phi_R(x)}$ for the right movers, and $L(x)\sim e^{i \sqrt{4\pi} \phi_L(x)}$ for the left movers
(up to factors of $1/\sqrt{2\pi}$) where $\phi_R(x)$ and  $\phi_L(x)$ are the chiral components of the non-chiral bosonic field  
$\Phi(x) = \phi_R(x) +\phi_L(x)$ that is formed along the barrier.
Introducing the dual bosonic field $\Theta(x)=\phi_R(x) - \phi_L(x)$, which satisfies the commutation relation $[\Theta(x),\p_x\Phi(x')]=-i\pi \delta(x-x')$,
the Hamiltonian density can be written as a sum ${\cal H} \sim \p_x\Phi(x) K_\Phi(x-x') \p_{x'}\Phi(x') + 
\p_x\Theta(x) K_\Theta(x-x') \p_{x'}\Theta(x')$.
The collective modes of the system are  $ \omega(q_x) = \pm \pi l^2 [\tilde K_{\Phi}(q_x) \tilde K_{\Theta}(q_x)]^{1/2}|q_{x}|$,
where $\tilde K_{\Phi}(q_x)$ and $\tilde K_{\Theta}(q_x)$ are Fourier transforms of the interaction kernels. 
The Fermi velocity in the case of long range interactions is momentum dependent.

To make some theoretical progress we assume here nevertheless that the interactions are short ranged. 
Such an assumption should not lead to significant changes on the physics of the system.
${\cal H}$ in this case takes the usual LL form with momentum independent Fermi velocity 
$\hbar v_F = (K_\Phi K_\Theta)^{1/2}$ and a LL constant 
$K  = (K_\Theta/K_\Phi)^{1/2}$. 

The Hamiltonian describing the system along the barrier in the the presence of tunneling 
$H_t =t_0 R^\dagger L +  h.c.$ is given
by
\begin{equation}
H =  \frac{\hbar v_F}{2}\int \rd x \left[(\p_x \Phi )^2 + (\p_x \Theta)^2\right]
  + 2\mu
\cos(\sqrt{4\pi K}\Phi (x))
\,,
\label{H_short_ranged}
\end{equation}
where $\mu=f(t_0,K)$. The $cos$ operator extends along the whole length of the barrier.
Only at the noninteracting point of $K=1$, is $\mu=t_0/2\pi$. The exact functional 
dependence $\mu=f(t_0,K)$, 
as shown by (Ref.~\onlinecite{kollar}), is dependent on microscopic details of the underlying system. 
We estimate the value of the LL parameter $K$ to be given by $K^2 \sim {(d^2/l^2)/\ln(L_x/l)}$, and differences in the value of $\mu$ from the free point value 
of $t_0/2\pi$ are expected.

Eq.~(\ref{H_short_ranged}) represents the well-known sine-Gordon model.
In this model the relevance of the electron tunneling processes 
between edges on opposite sides of the barrier 
depend on the value of the parameter $K$. This parameter on the other hand depends on the value of the filling 
fraction of the QH liquids on the sides of the barrier, the interaction strength along the QH line junction, and the interaction 
strength between charges of counterpropagating edges on each QH subsystem on each side of the barrier. 
From previous calculations \cite{papa} the interactions  across the Hall bar and the interactions along the line junction 
have the opposite effect on the relevance of the tunneling processes. In general there can be geometries in which these interactions can be tuned 
to cancel each others effect. 
It is assumed however that the Coulomb interactions between charges located on the edges on the top and bottom of the Hall bar in 
systems examined by Kang {\it et al} \cite{kang}, 
due to their spatial separation, are small thus can be neglected in the following.

The sine-Gordon model obtained above is a well-known model in quantum field theory and many of its properties are known. 
This model has a discrete symmetry $\Phi \rightarrow \Phi + 2\pi n/\beta$, where $\beta^2=4\pi K$. 
At $\beta^2/8\pi <1$  in the ground state the symmetry is spontaneously broken and the spectrum becomes massive. 
In the following we absorb the factor $1/8\pi$ in the definition for $\beta$.
In the region $1/2 < \beta^2 < 1$, known as the "repulsive" regime,
the spectrum is composed of topologically charged particles called solitons and antisolitons. 
In the region $0 < \beta^2 < 1/2$, known as the "attractive" regime, reached in the presence of forward scattering 
interaction, the spectrum is enriched by the their bound 
states, called  breathers, the number of which increases with decreasing $\beta$. 
The mass spectrum of the model is 
\bea
M_n = 2M_{\rm s} \sin\left(n \frac{\pi}{2} \frac{\beta^2}{1-\beta^2}\right)
\, , \,
n=1,\ldots,\left[\frac{1-\beta^2}{\beta^2} \right] \,,
\label{mass_spectrum}
\eea
where $n$ is integer and  $[\cdot]$ represents the integer part of a real number.
With $M_n$ we denote the mass of the $n$-th breather, whereas with $M_s$ we denote the mass of the soliton. 
The soliton and antisoliton have equal mass.

The scaling dimensions of the $cos$ operator are $d=K/2$ and at $K \le 2$ the QHLJ turns into an insulator along its length 
when the chemical potential for topological particles falls withing the gap. In this case the QH effect is established along 
the whole sample and current flows along its perimeter. Therefore at zero temperature 
one would expect the height of the zero-bias conductance peak to be $e^2/h$. At finite temperatures however
there will be some particles in the spectrum even at chemical potential  
below the mass gap. As a result even when the chemical potential falls within the gap, current will partly be carried along the junction. 
When the chemical potential falls outside the gap the QHLJ becomes an 1D conductor and at low voltages and low temperatures 
the conductance across the barrier vanishes.

The mass of the physical particles of the SG model, in the context of the QHLJ systems with interedge tunneling present, 
constitutes half of the gap that is measured in the experiments of Ref.~\onlinecite{kang}. It depends on the microscopic parameters of the system,
the exact value of the Fermi velocity, coupling constant and tunneling amplitude 
as follows
\cite{zam_95} 
\bea
\frac{M_s\Lambda}{\hbar v_F}= \frac{2}{\sqrt{\pi}}
\frac{\Gamma\left(\frac{\beta^2}{2(1-\beta^2)}\right)}{\Gamma\left(\frac{1}{2}\frac{1}{1-\beta^2}\right)} 
\left[\frac{\mu \Lambda}{\hbar v_F}  \frac{\pi \Gamma(1-\beta^2)}{\Gamma(\beta^2)}\right]^{\frac{1}{[2(1-\beta^2)]}} \, ,
\label{mass}
\eea
where $\Lambda$ is a short distance cutoff of the field theory. These quantities however are very hard to estimate  with accuracy
in simplified microscopic models. This is the reason that we try to present here indirect ways of estimating these quantities experimentally.

\section{Experimental possibilities for estimation of $\beta^2$, $v_F$}
\label{drude}

As it was discussed in Refs.~\onlinecite{haldane,papa3}, even though the SG model has a massive spectrum, 
massless modes can be induced in 
the presence of external fields that couple to the charges of the topological excitations. 
We show below how the parameters $\beta$ and $v_F$ can be extracted using the thermodynamic Bethe ansatz method.
In QHLJ that we are considering here, we discuss two situations where the gap suppression occurs.

I. The first one is the case when the filling fractions on the 2DEG subsystems is $\nu=1$. In this 
case   
such a field can be some linear external charge $Q(x)$ that couples to the charge of the 1D non-chiral LL that extends 
along the separating barrier, and can be represented as
\bea
H_Q = \int \rd x \rd x' Q(x)V(x-x')\rho(x') \quad .
\eea
$V(x-x')$ is an interaction kernel. Just as in the case of Eq.~(\ref{H_short_ranged}) 
we assume here that the interactions 
are short ranged due to the presence of screening gates or other metals in the vicinity of the barrier.
The total charge density as usual in the bosonization approach is represented by
\bea
\rho(x) = \rho_0 + \frac{1}{\sqrt{\pi}} \p_x \Phi(x) \quad ,
\eea
where the constant charge density $\rho_0 = k_F/\pi$ is usually neutralized by the positive background charge.

II. The other possibility that we will consider below arises at bulk fillings $\nu \sim 2$ of the 2DEG subsystems. 
In this case the electron spin is not frozen as in the previous case. The spin degree of freedom plays an important 
role on the dynamics of the edges. 
The sine-Gordon model can arise here either in the presence of antiferromagnetic interactions 
or in the presence of ferromagnetic type interactions with magnetic anisotropy (Ref.~\onlinecite{kim}), 
as we elaborate below.
The external field that can drive the transition to the gapless phase can be the magnetic field 
(that couples with the spin current)
\be
H_{\rm Zeem} = -\mu_B g B (J^z+\bar{J^z})  = \frac{\mu_B g_e B }{\sqrt{\pi}} \p_x \Phi_s(x)\quad ,
\label{Zeeman}
\ee
where $J^z+\bar{J^z} = -\p_x\Phi_s/\sqrt{\pi}$ was used. In Eq.~(\ref{Zeeman}) 
$g_e$ is the gyromagnetic ratio of the electron and $\mu_B$ is the Bohr magneton.

For the field that couples with the gradient of the field $\p_x \Phi$, whether we will be talking for the first or second case, we will use the notation 
$H$, where we will mean either $H\sim QV$ or $H\sim \mu_B g_e B$.

The presence of the $H$ field in the Hamiltonian, has important consequences in the physics of the system.
In the absence of $H$, the ground state of the system is reached at the trivial, the static and 
constant in space field configurations $\Phi(x) = 2\pi n/\beta$, being the discrete degenerate minima of the potential energy.  
The presence, on the other hand, of $H\p_x\Phi(x)$ and the harmonic term in the Hamiltonian as optimal require the new field 
configuration be $\Phi(x)\sim Hx$. These competing tendencies are resolved at a critical value $H_c$ of $H$, above which the system 
is characterized by a finite value $\left < \p_x \Phi(x)\right> \neq 0$ 
(or finite density of solitons).  In the new phase the field $\Phi$ generally follows the $Hx$ line, but such as to minimize 
also the potential energy $\cos \beta\Phi(x)$, leading to finite regions of peaks in the charge density.
This is the commensurate-incommensurate transition, first discussed in Ref.\onlinecite{pokrovsky}. 

On the quantum level, $H$ serves as chemical potential for the topological particles. In its presence the solitons lower 
their energy, and above a critical value of the chemical potential, they will tend to 
proliferate in the ground state. 
These particles however interact with each other, leading to a Fermi-sea-like distribution in the ground state. 
Gapless modes would consist of excitations above this Fermi sea.
Our aim here is to find the value of the parameters of the SG model, $\beta$ and $v_F$, discussed in the 
previous sections.
With the thermodynamic Bethe ansatz, we can calculate the values of the LL parameter $K$ of this metallic phase, as well as the 
corresponding Fermi velocity $v_F$, as functions of both, chemical potential as well as the SG parameters $\beta$ and $v_{F}$.
$K$ and $v_F$  at large values of the chemical potential relative to the mass gap approach respectively values
of $\beta^2$ and $v_{F}$ of the insulating phase. 
However, we show that already at $H\sim 1.5 (\Delta/2)$
these quantities are practically to within 5\% range 
of the values of $\beta^2$ and $v_{F}$ of the SG model.   

From the practical point of view, in the experimental studies, one way of getting
access to the LL parameters in the incommensurate (I) metallic phase is by measuring the Drude weight, which is given
by the product $v^I_F K$. 
Another way to access a different combination of $v^I_F$
and $K$ is by measuring the charge susceptibility, which is given by their
ratio $\chi \sim K/v^I_F$.

\begin{figure}
\unitlength=1mm
\begin{picture}(80,53)
\put(-3,4){\includegraphics[height=50mm]{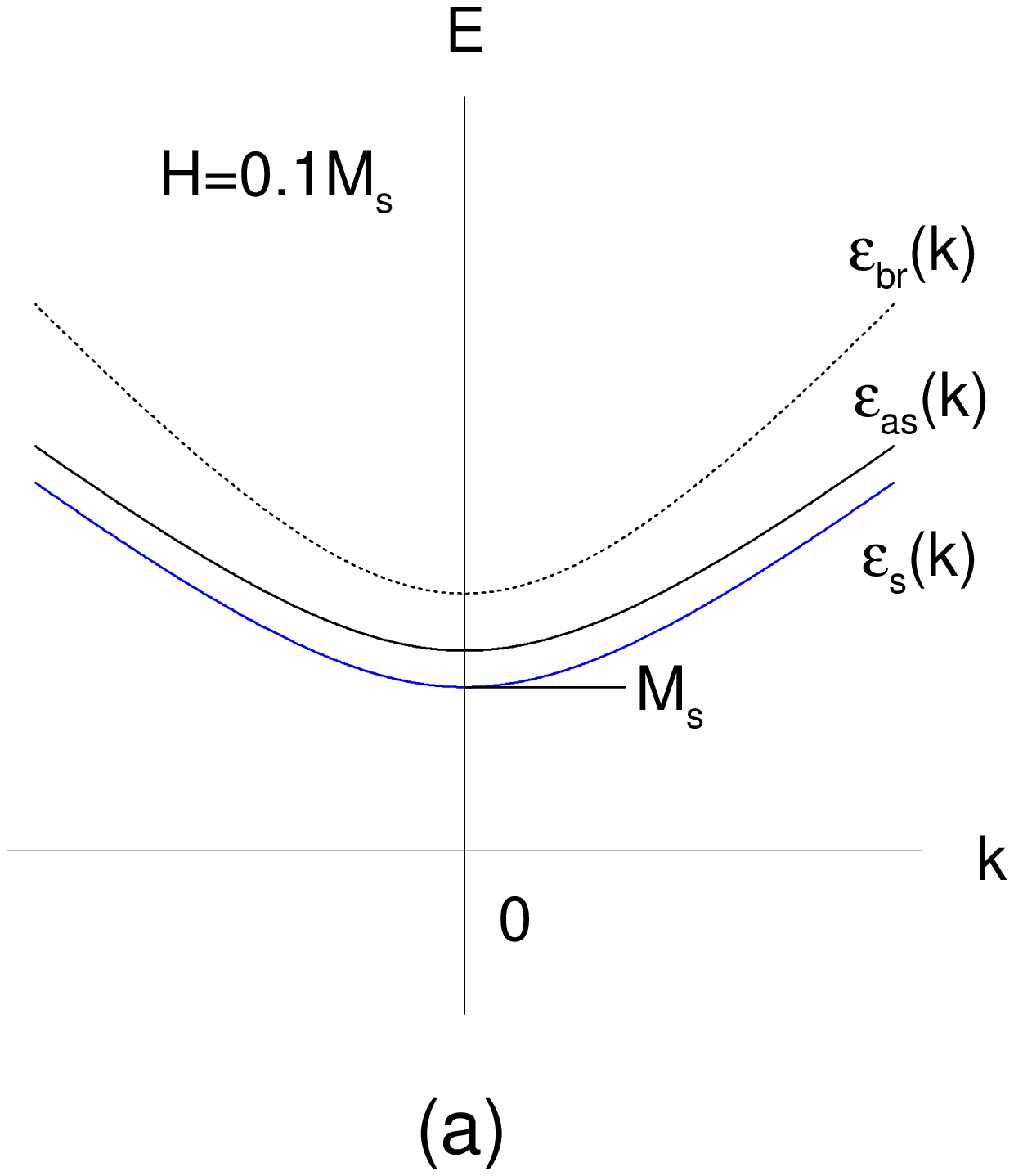}}
\put(44,4){\includegraphics[height=50mm]{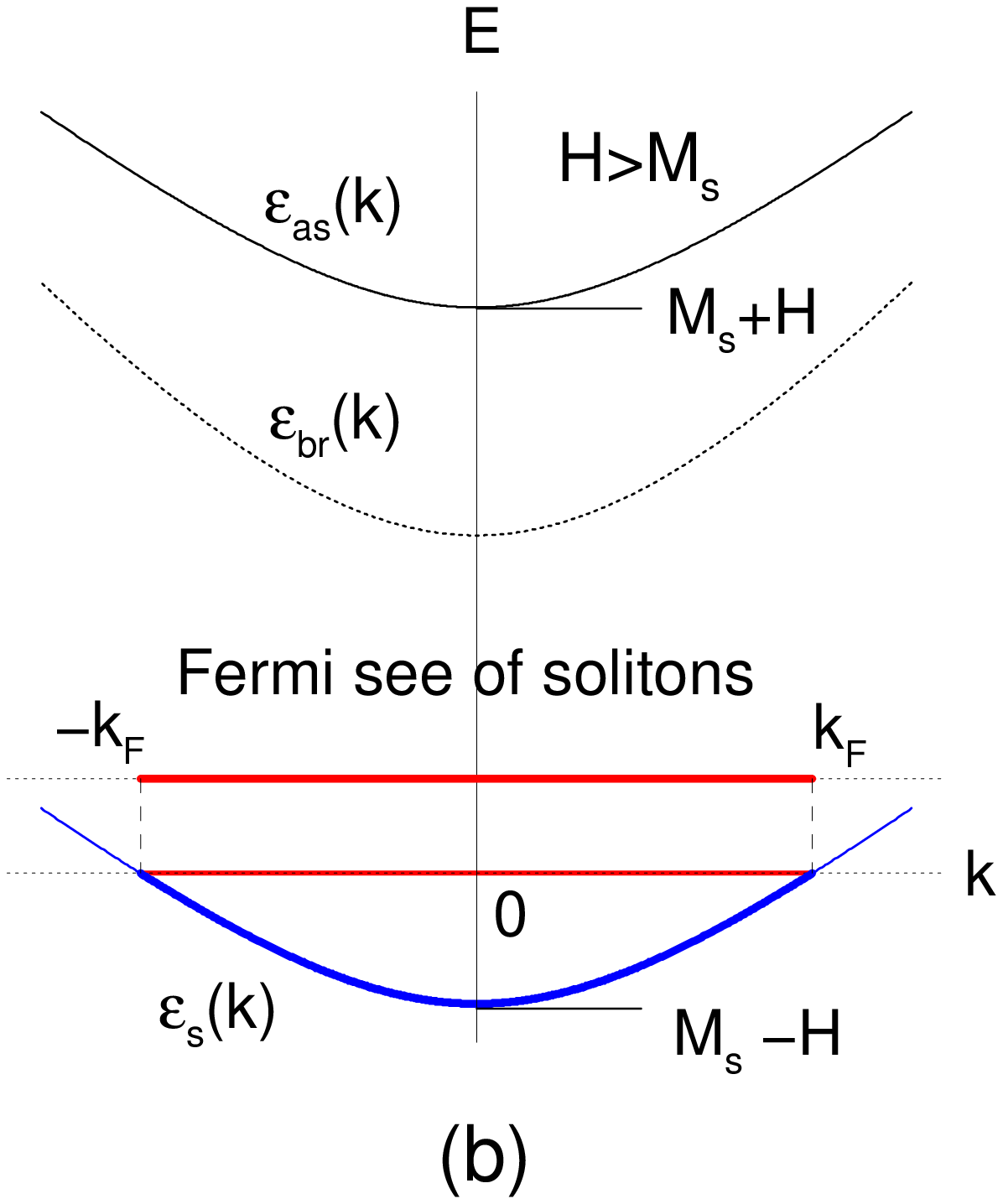}}
\end{picture}
\caption{ On the left, the energy spectrum of the SG model in the presence of a 
small external 
field $H$, here $H/M_s=0.1$, to allow for a small split in the spectrum, is shown. 
In the absence of an external field and at $T=0$ the soliton and antisoliton energies coincide, whereas 
the breathers energy is slightly larger [see Eq.~(\ref{mass_spectrum})]. 
For this representation the temperature was taken to be $T/M_s=0.01$.  On the right, the spectrum 
in the presence of an external field that exceeds the soliton mass, is shown. 
For $H>M_s$ the soliton gap closes and in the ground state a soliton condensate forms. 
Due to the opposite sign of their topological charge the energy of the antisolitons increases.
The system is at $T=0$ or very small $T$, compared to $M_s$, a solitonic Luttinger liquid.
The graphs are obtained from the numerical 
solutions of the defining integral equations.}
\label{SG_LL_schemat}
\end{figure}

In the following first we discuss the applicability of the LL picture at finite temperatures.

\subsection{
Luttinger Liquid picture of sine-Gordon model in the presence of external fields: Finite temperatures}

Before discussing the question of the Drude weight and charge susceptibility in the metallic phase of the SG model we first discuss 
the applicability of the LL picture at finite temperatures.
The case that we are discussing, the LL is obtained by linearizing the spectrum around the two Fermi points of the soliton sea.
This approximation works at its best at $T=0$, and its accuracy decreases with increasing temperature.
The experiments from the other side are never performed at zero temperatures. It is useful therefore to examine its applicability
at finite $T$.

With increasing temperatures the particles start filling states away from the Fermi surface and at relatively higher temperatures 
the breathers and antisoliton branches start filling too. As a result the low energy excitations are not of a particle-hole 
type close to the Fermi surface any more.
Here, 
the applicability of the LL picture is examined by studying the dependence of the susceptibility $\chi$ on $T$. 
Its divergence at a certain critical value of the external field notifies a transition 
from a gapped system to a gapless one with sharp Fermi surface. 
At finite temperatures there will be some particles in the spectrum even at external fields below the mass gap, as a result the
divergence turns to a finite height peak which decreases continuously with temperature.

To obtain the dependence of the susceptibility on the external field $H$  and temperature $T$ exactly,
we use the thermodynamic Bethe Ansatz (TBA), a summary of which is contained in the Appendix.

Here we examine the SG model only at the reflectionless points $\beta^2=1/n$, 
where the particle scattering is diagonal
(except the asymptotic case $H/T\gg 1 $, which is applicable for any $\beta$).
In the following the calculations are performed in rapidity space $\theta$ which parametrizes
the energy and momenta as follows
\bea
E_a(\theta) = m_a\cosh\theta \quad , \quad P_a(\theta)=\frac{m_a}{v_F}\sinh\theta \quad ,
\eea
preserving thus the relativistic dispersion relation
\bea
E_a^2(P_a)= v_F^2 P_a^2 + m_a^2 \quad .
\label{quadratic_spectrum}
\eea
$m_a$ are particles' masses  and $v_F$ their velocity.
Before proceeding any further a few words are in order on the equations for the spectral functions obtained with the 
Bethe ansatz technique that we will use in the following. 
The Bethe ansatz equations are obtained from the set of periodic boundary conditions set on the phases 
accumulated 
by particles with different momentum as they travel around the circumference of a circle of length $L$ in 
the presence of the other particles.
In these equations the interactions are reflected in the 
phase collected as two particles interchange positions. 
First of all, it is a general fact that no two particles in integrable theories can occupy the same state (for then the 
Bethe wave function that describes the overall system vanishes). 
Therefore in these models, in the ground state, the particles are distributed on a Fermi sea. 
The energy spectrum is determined by  a complete system of coupled integral equations
\begin{equation}
m_a \cosh \theta -H_a q_a = \epsilon_a(\theta) + T\sum_{b}
K_{ab}*\ln\left[1+e^{-\epsilon_b/T} \right]\, ,
\label{set-of-dress-e1}
\end{equation}
where $(*)$ is a short notation for the convolution of $K_{ab}(\theta-\theta')$ with 
$\ln\left[1+\exp\{-\epsilon_b(\theta')/T\} \right]$.
In Eq.~(\ref{set-of-dress-e1}) the subscripts $a$, $b$ stand for $s$, $\bar{s}$, for soliton, antisoliton, and $n$ for 
the $n$-th breather, respectively.  
$H_a$ is a chemical potential for the particles of the spectrum.
If interactions are neglected
the spectral function agrees with the bare one [left-hand side of Eq.~(\ref{set-of-dress-e1})].
Their topological charges are 
$q_s=-q_{\bar s}= 1$ and $q_n=0$. Due to the sign difference of the charges of solitons and antisolitons 
in the presence of a chemical potential their spectra become different.  Particularly at $H >  M_s$, the energy spectrum will 
contain an interval of states of negative energy. 
At $T=0$ and $H > M_s$, due to the fact that only one of the spectra 
takes negative values between the points $\pm B$ (the Fermi points), the equations (\ref{set-of-dress-e1}) simplify. 
The term involving $K_{ab}$ on the right-hand side of 
Eq.~(\ref{set-of-dress-e1}) reduces to $[K_{ss}*\epsilon_s](\theta)$ in the interval $(-B,B)$. The Fermi points are 
defined self consistently from  
$\epsilon_s(\pm B)=0$.
The Kernels $K_{ab}(\theta)$ are obtained from the scattering matrices by taking
\bea
K_{ab}(\theta)= - \frac{i}{2\pi}\frac{\rd}{\rd \theta} \ln S_{ab}(\theta) \quad .
\eea
The scattering matrices encode all the information about the particle interactions.
For the sine-Gordon model they are known (Refs.~\onlinecite{zam_77,zam_79,klassen,fendley_S_W}), and for completeness 
we give them in the following
\bea
K_{nl} & = & \int_{-\infty}^{\infty} \frac{\rd \omega e^{i\omega\theta}}{2\pi}
\left[
\delta_{nl} -
\right.
\nonumber\\[3mm]
&&
\left.
-2\frac{ \cosh{\left(\xi\frac{\pi}{2}\omega\right)}\cosh[(1 -n\xi)\frac{\pi}{2}\omega]
\sinh{\left(l\xi\frac{\pi}{2}\omega\right)}}{\cosh{\left(\frac{\pi}{2}\omega\right)}
\sinh{\left(\xi\frac{\pi}{2}\omega\right)} }\right] \ ,
\nonumber\\[3mm]
K_{n,\pm} & = & - \int_{-\infty}^{\infty} \frac{\rd \omega e^{i\omega\theta}}{2\pi}
\frac{ \cosh{\left(\xi\frac{\pi}{2}\omega\right)}\sinh{\left(n\xi\frac{\pi}{2}\omega\right)}}{
\cosh{\left(\frac{\pi}{2}\omega\right)}\sinh{\left(\xi\frac{\pi}{2}\omega\right)} } \ ,
\\[3mm] \nonumber
K_{\pm,\pm}  & = & K_{\pm,\mp} = -\int_{-\infty}^{\infty} \frac{\rd \omega e^{i\omega\theta}}{2\pi}
 \frac{ \sinh{\left[(1-\xi)\frac{\pi}{2}\omega\right]}}{2\cosh{\left(\frac{\pi}{2}\omega\right)}
\sinh{\left(\xi\frac{\pi}{2}\omega\right)}}  \, ,
\\[3mm]\nonumber
n,l & = & 1,\ldots,\frac{1}{\xi}-1;\quad n\ge l \quad, \quad \xi = \frac{\beta^2}{1-\beta^2} \quad .
\eea
In Eqs.~(\ref{T_0_spectr}-\ref{density}) below, the minus sign is absorbed in the Kernel.

The Kernels can be calculated easily for the cases when $\beta^2 =1/n$, for $n$ integer. 
In the Appendix we give the Kernels for the $n=3$ and $n=5$ when the spectrum contains just one and three breathers, respectively.

The integral equations for the energies of the particles Eq.~(\ref{set-of-dress-e1}) can be trivially solved at the free fermion Luther-Emery point 
$\beta^2=1/2$.   
At all other values of $\beta^2$ they can be solved numerically.  
The number of coupled integral equations increases increases with decreasing value of the coupling constant $\beta^2$.

\begin{figure}
\unitlength=1mm
\begin{picture}(80,40)
\put(-1.5,2){\includegraphics[height=41mm]{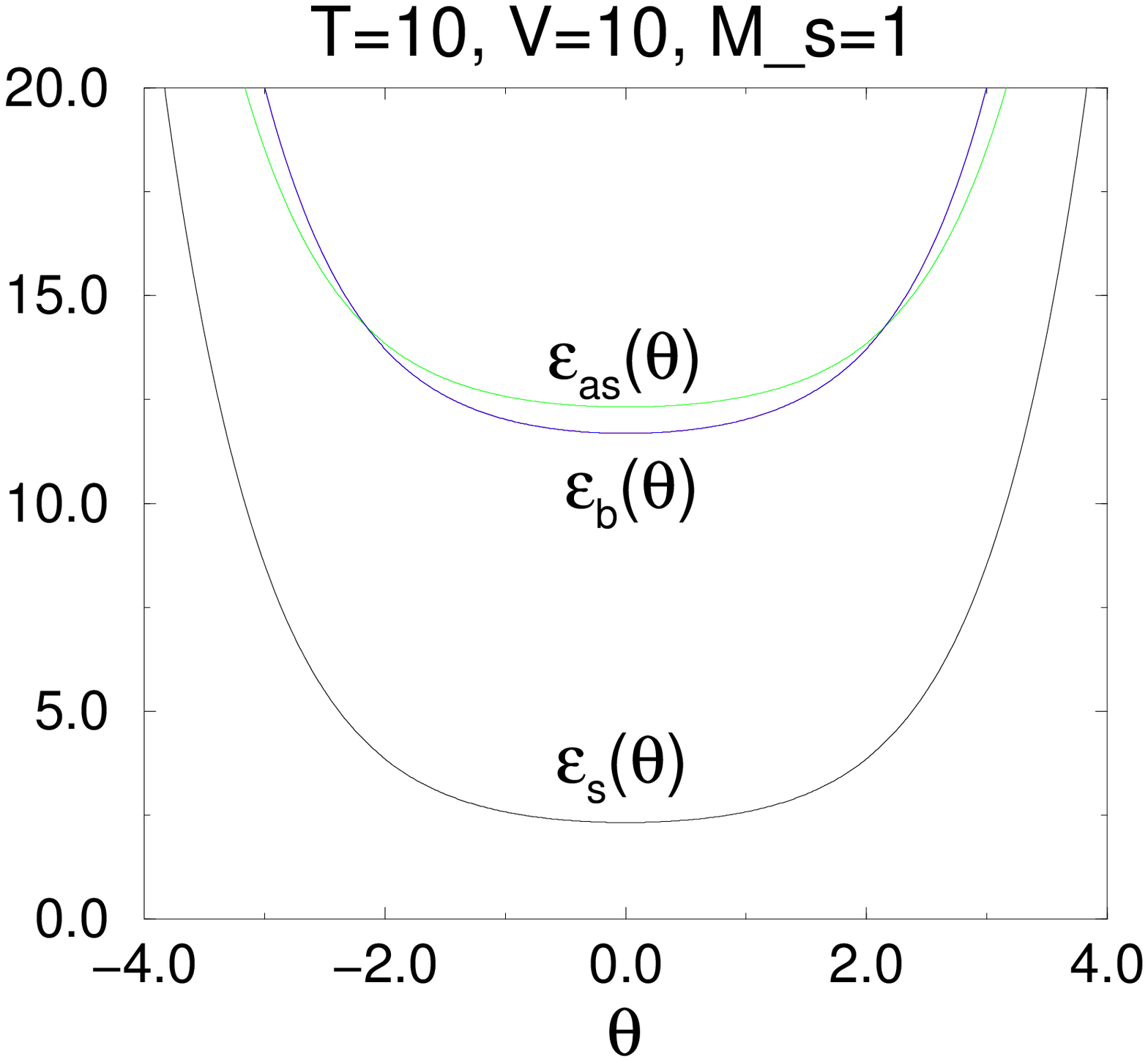}}
\put(40,1.7){\includegraphics[height=41.2mm]{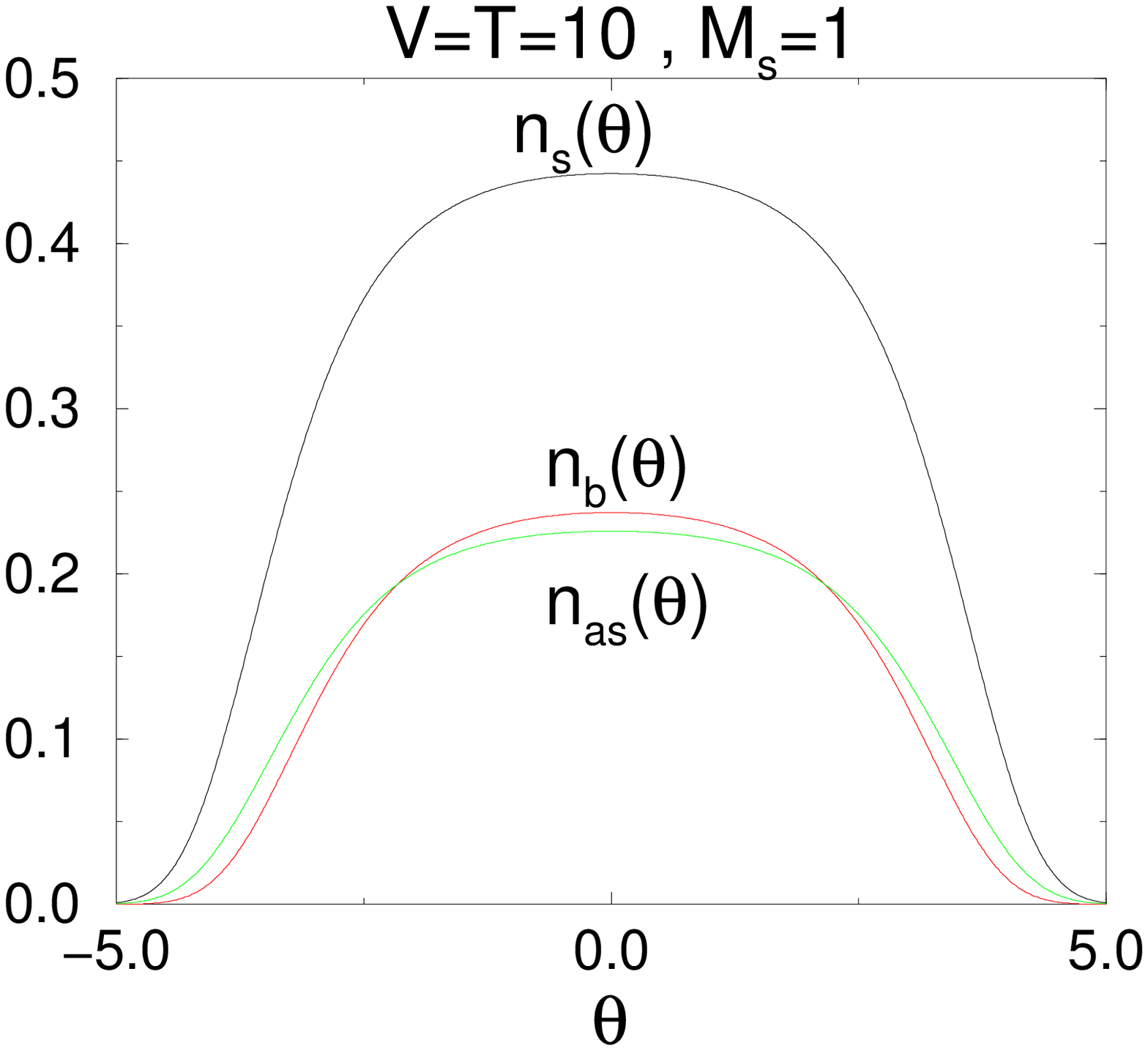}}
\end{picture}
\caption{ On the left panel the energy spectra of solitons, breathers and antisolitons, respectively, in
rapidity space are presented.
On the right panel the respective distribution functions are presented. The particles behave
as fermions for $\theta\rightarrow 0$, no two particles of the same kind with the same rapidity can exist. 
These particles are neither fermions nor bosons. The phase accumulated by interchanging two particles 
depends on the difference of their rapidities.  At zero
temperature and in the presence of field $V> 2M_{\rm s}$ the antisoloton and breather branches remain unpopulated
whereas the solitons reside in the interval $(-B,B)$ with a step function distribution in rapidity space.
$B$ is defined by the equation $\epsilon_s(\pm B)=0$ and Eq.~(\ref{set-of-dress-e1}) for $T=0$.
At $T\gg M_s$, which is the case presented here, the antisoliton and breather particles are also present.
}
\label{Massive_SG_spectra}
\end{figure}

In Fig.~\ref{Massive_SG_spectra} (left) we give plots of the energy spectra at $\beta^2=1/3$ and
 $k_BT=eV=10M_s$, where $V=2H$. In Fig.~\ref{Massive_SG_spectra} (right) also the particles' distribution functions are shown. 
The fact that no two particles with same momenta occupy the same state is an well-known feature of the integrable models, 
resembling this way the fermionic statistics. In fact they have a mixed  or anyonic 
statistics since their scattering matrix depends on the difference of their rapidities.  These particles are neither fermions nor bosons.
The phase accumulated by interchanging these particles will depend on the difference of their rapidities.
Notice, for instance,   that interchanging particles with $\theta_1-\theta_2\rightarrow +\infty$, results in $S_0(\theta_1-\theta_2)=1$, 
thus resembling bosons.
 
From BA, for the susceptibility we have
\bea
\chi = -\frac{\p^2 f(\beta^2,H,T)}{\p H^2} \quad ,
\label{TBA_susc}
\eea
where $f$ is the free energy of the system in the presence of a chemical potential. In TBA it is given by
\bea
f= - T \sum_a m_a \int_{-\infty}^{+\infty} \frac{\rd \theta}{2\pi}\cosh \theta
\ln\left[1+e^{-\epsilon_a(\theta)/T}\right] \, ,
\label{TBA_free_en}
\eea
where $\epsilon_a(\theta)$ is the energy of the particles in the presence of interactions.

\subsubsection{Asymptotic limit $H/M_s \gg 1$ of charge susceptibility at $T\neq 0$}

In the limit of large values of chemical potential above the mass gap, 
the functional dependence of the free energy (and therefore of the  susceptibility) on $\beta^2$, and $H$, can be obtained analytically. 

In the  $H/M_s\gg 1$ limit and temperatures of the order of the soliton mass, $T\sim M_s$, one can assume that the 
spectrum contains only solitons. Breathers and antisolitons are also present, but as can be inferred from 
Figs.~\ref{SG_LL_schemat}, \ref{Massive_SG_spectra}, their density is small
due to high energies required to fill these branches. 
The formula for the free energy Eq.~(\ref{set-of-dress-e1})  in this limit 
simplifies to a single integral, accounting simply for the energy of the solitons.
Moreover, due to the large density of states available to solitons at higher rapidities (reached at high $H$), 
the main contribution in the free energy comes from the region $\cosh\theta \sim H/M_s - 1$. Here we can  
approximate $\cosh\theta \sim e^\theta/2$, and rewrite the equation for the soliton energy
\bea
\epsilon_s(\theta) + T K_{ss} *
\ln\left[1+ e^{-\epsilon_s/T}\right] (\theta) = \frac{M_s e^\theta}{2} - H \,.
\label{asympt_eps}
\eea
In this limit it is assumed that all solitons have $v_F^I \approx v_F$, and the TBA equations and
$\epsilon_s(\theta)$,  resemble the massless case of Fendley {\it et al.}
\cite{FLS}. This $\epsilon_s(\theta)$ differs from the original one at the low energies. 
It turns out that $\epsilon_s(-\infty)$, will enter the expression for the functional dependence of $f$ on $H$, $\beta^2$,
and $T$.

Using Eq.~(\ref{asympt_eps}), we can split the free energy into two parts. After technical steps\cite{saleur},
we find that at $H/T \gg 1$, the first one $f_1$ converges to the $H$ independent 
constant $f_1 = -\pi T^2/12$. 
In fact the leading order terms in $f_1$ are
\begin{equation}
f_1 = -\frac{\pi T^2}{12} + \frac{T^2}{2\pi}e^{{\epsilon_s(-\infty)}/{T}}  
\hspace{-0mm} - 
\frac{T \epsilon_s(-\infty)}{4\pi}e^{{\epsilon_s(-\infty)}/{T}} +\ldots \, .
\end{equation}
The second part, with $H$ contained in leading order, 
turns out to converge to $f_2=-\beta^2H^2/\pi$. This is the part that we are interested here.
More precisely $f_2$ is given by 
\bea
f_2 
& = & - \frac{HT}{2\pi}\ln\left[1 + e^{-\epsilon_s(-\infty)/T}\right] \, ,
\eea
where $\epsilon_s(-\infty)$ can be found from Eq.~(\ref{asympt_eps}).

For the range of temperatures $T\le M_s$ and chemical potentials  $H \gg M_s$  
that we are interested here,  we find
\bea
\epsilon_s(-\infty)  
&=& 
-\frac{H}{(1/2\beta^2)} + T (1-2\beta^2) e^{-2\frac{\beta^2 H}{T}} \; .
\eea
The interactions renormalize $\epsilon_s(- \infty)$ up or down changing it from 
$\epsilon_s(- \infty)=-H$ (the bare energy) of the first order of approximation, 
the noninetarcting limit, to $\epsilon_s(- \infty)=-2\beta^2 H$ when interactions are present ("dressed" energy).
The renormalization occurs due to interactions with other particles, present in the presence of large $H$.
Therefore the asymptotic behavior of the free energy for the above case  of $H\gg M_s \ge  T$ will be
\begin{equation}
f(H) = 
-\frac{\beta^2 H^2}{\pi} - \frac{\pi T^2}{12} - \left[\frac{\beta^2T H}{2\pi} 
- \frac{T^2}{2\pi} \right] e^{-\frac{2\beta^2 H}{T}} +\ldots\, ,
\label{free_en_asymp}
\end{equation}
which leads to a constant susceptibility up to higher temperatures. 
At $H\gg M_s \ge  T$ (putting back also $v_{F}$) the charge susceptibility  
approaches the constant value of $\chi \rightarrow 2\beta^2/\pi v_F$
\bea
\chi(H/\Delta\gg 1) = \frac{2\beta^2}{\pi v_{F}} \quad .
\eea
This corresponds to the limit of large density of solitons.

\subsubsection{Numerical results for the charge susceptibility at finite temperatures}

At values of the external field on the order of the gap or smaller and at finite temperatures
the breathers and antisolitons give significant contributions to the thermodynamics of the system.
For $\beta^2=1/2$ for the free energy in the region $H < M_s$ we find
\be
f=\frac{2\Gamma(3/2)}{\pi^{3/2}} (M_s T)\sum_{n=1}^{\infty}\frac{(-1)^n}{n} 
2 \cosh\left(\frac{n H}{T}\right) K_1\left(\frac{n M_s}{T}\right),
\ee
where $ K_1$ is the modified Bessel function.

\begin{figure}
\unitlength=1mm
\begin{picture}(80,48)
\put(11,2){\includegraphics[height=45mm]{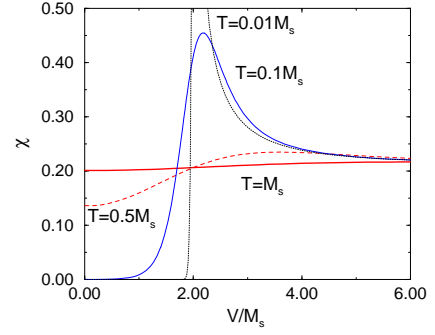}}
\end{picture}
\caption{The charge susceptibility of the SG model as function of the external field $V=2H$, where $H$ is 
the chemical field for solitons and antisolitons, is presented.
The Luttinger liquid description seems to be not useful even at temperatures $T=M_s$. 
The asymptotic behavior $H/T\gg 1$ found numerically agrees with the analytical 
result $\chi=2\beta^2/\pi$, of Eq.~(\ref{free_en_asymp}). Here $\beta^2=1/3$.}
\label{Massive_SG_figures}
\end{figure}

\begin{figure}
\unitlength=1mm
\begin{picture}(80,50)
\put(11,2){\includegraphics[height=47mm]{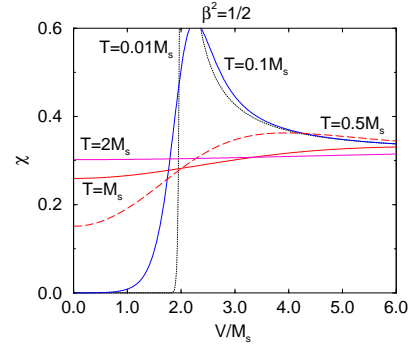}}
\end{picture}
\caption{The charge susceptibility of the SG model as function of the external field $V=2H$ for the case of $\beta^2=1/2$, 
is presented.
The Luttinger liquid description in this case starts to break down at higher temperatures 
compared with the case of $\beta^2=1/3$.  }
\label{Massive_SG_figures_2}
\end{figure}

Plots of the susceptibility in the whole range of the external field and for various temperatures
compared to the gap are given in Figs.~\ref{Massive_SG_figures} and \ref{Massive_SG_figures_2} for values of the coupling constant $\beta^2=1/3$ and $\beta^2=1/2$, respectively. In both cases at $T=0$ a square 
root singularity occurs at $V/\Delta=1$. This is the result of the square root dependence 
of the Fermi velocity on the external field in the metallic phase at $V/\Delta>1$.
At finite temperatures the divergence  in $\chi$ at the gap crossing is suppressed. 
At higher values of the field $H$ however the lines for $\chi$ obtained at various 
temperatures merge together.

The more breathers in the spectrum, the faster the LL description breaks-down. 
Since in the case of Fig.~\ref{Massive_SG_figures_2} in the spectrum there are no breathers the LL description is expected to be a 
good approximation till higher temperatures than the case of smaller $\beta^2$.
The comparison between Fig.~\ref{Massive_SG_figures} and Fig.~\ref{Massive_SG_figures_2} makes clear this intuitive expectation.

At values of the chemical potential reasonably close to the mass gap 
an approximate value of the temperature for the validity of the LL description of the model at 
finite temperatures for the case of $\beta^2=1/3$ would be $T\sim 0.1M_s \sim 0.75K$. In cases 
of stronger interactions this temperature is expected to become progressively lower.

\subsection{Drude weight and charge-charge correlation function in I phase of the sine-Gordon 
model at $T=0$}

Here we calculate the Drude weight and the charge susceptibility for the gapless modes above the
soliton condensate when the system can be described as a LL. 
At low energies and at $T=0$ the excitations of the system can be described by the LL action 
\bea
S=\frac{\hbar}{2 K} \int \rd t \rd x \bigl[\frac{1}{v^I_F}\left(\p_t \Phi \right)^2 - 
v^I_F \left(\p_x \Phi\right)^2\bigr] \quad,
\eea
where $v^I_F$ is the Fermi velocity in the incommensurate phase and $K$ the LL parameter. 
This kind of LL, which is obtained by linearizing the spectrum around the two Fermi points of the soliton sea, has parameters that 
clearly should depend on the values of $H$ and the coupling constant $\beta^2$ of the underlining SG model.
For the relationship of $v^I_F$ and $K$ to $\beta^2$ and $H$ we use the exact solution of SG.

The Drude weight and charge susceptibility gives an way to experimentally  measure the combinations
$v^I_FK$ and $K/v^I_F$. They are both functions of $\beta^2$ and $H$ and we examine in this 
section this dependence. We show that the Fermi velocity in the I phase converges to $v_F$ 
very quickly with increasing $H$. $v_F$ is the Fermi velocity contained in Eq.~(\ref{H_short_ranged}) 
the velocity prior to the gap opening.
The latter parameter is the one that is needed for the 
evaluation of the gap in Eq.~(\ref{mass}). The other remaining parameter needed to find $M_s$ is the value 
of the coupling constant $\beta^2$. As it was shown in Ref.~\onlinecite{papa3} the LL parameter $K$ converges fast 
to its asymptotic value $K\rightarrow \beta^2/(4\pi)$ as $H$ increases above $H_c=M_s$.

The optical absorption is proportional to the real part of the optical conductivity. The real part of the optical conductivity 
on the other hand is related to the imaginary part of the time ordered current-current correlation function by
\bea
\Re{\rm e}\{\sigma(\omega,q_x)\} = \Im{\rm m}\Big[\frac{\Pi(i\omega\rightarrow \omega +i 0^+,q_x)}{\omega}\Big]
\quad ,
\label{opt_cond}
\eea
where the Fourier transform of the time ordered current-current correlation function
is given by
\bea
\Pi(i{\omega},q_x) = \int \rd x \rd \tau e^{i{\omega} \tau} e^{-iq_x x} T_\tau \left< j(x,\tau) j(0,0)\right>
\;.
\eea
Since the electrical current can be expressed in terms of the right and left chiral currents  by
$j(x,t)= v^I_F [J(z) - \bar{J}(\bar{z})]=(i/\sqrt{\pi})\p_\tau \Phi(x,\tau)$, Eq.~(\ref{opt_cond}) takes the form
\bea
\label{conductivity}
&& \Re{\rm e}\{\sigma(q_x,\omega)\} = \Im{\rm m}\Big\{-\frac{1}{\pi\omega}\int \rd \tau \rd x
\Big. 
\\[3mm]\nonumber
&& \hspace{15mm}\Big. 
T_\tau \left<\p_\tau \Phi(x,\tau) \p_\tau
\Phi(0,0)\right>e^{i{\omega} \tau} e^{-iq_x x}   \Big\}
\quad .
\eea
The units of the conductivity $(e^2/h)$ are not written out explicitly. 
Eq.~(\ref{conductivity}) leads to 
\be
\Re{\rm e}\{\sigma(q_x,\omega)\}  =  - \frac{v^I_F K}{\pi} \Im{\rm m}
\Big\{
\frac{\omega + i 0^{+} }{(\omega +i 0^{+} )^2 - (v^I_F)^2 q_x^2 } \Big\}
\ee
and 
\bea
\Re{\rm e}\Big\{\sigma(q_x \rightarrow 0,\omega \rightarrow 0)\Big\} = v^I_F K \delta(\omega)
\quad .
\eea
The Drude weight will be $D=v^I_F K$, where $K$ is the Luttinger liquid parameter.

\begin{figure}
\unitlength=1mm
\begin{picture}(80,45)
\put(8,5){\includegraphics[height=42mm]{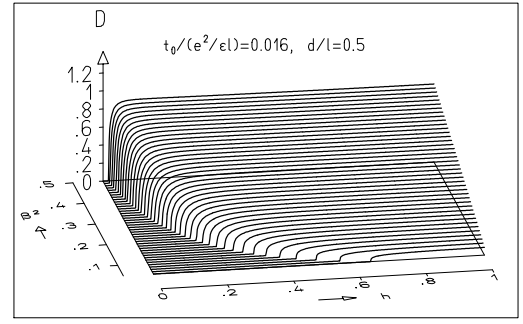}}
\end{picture}
\caption{The Drude weight in the incommensurate (metallic) phase of the SG model as function of $\beta^2$ and 
external field is shown.}
\label{modes}
\end{figure}

The external field couples to the charge 
$\rho(x,t)=J(z) + \bar{J}({\bar z}) = -\p_x \Phi(x,t)/\sqrt{\pi}$. 
The charge susceptibility will be given by a similar equation to Eq.~(\ref{conductivity}),
now through the time ordered charge-charge correlation function
\bea
\label{susceptibility}
\chi(q_x,\omega) &=& \frac{1}{\pi}\int \rd \tau \rd x
\\[3mm]\nonumber
&&T_\tau \left<\p_x \Phi(x,\tau) \p_x
\Phi(0,0)\right>e^{i{\omega} \tau} e^{-iq_x x}\big|_{i{\omega} \to \omega + i0^+}   
\; ,
\\[3mm]
&=&\frac{v^I_F K}{\pi } \frac{q_x^2}{{(\omega  + i 0^+)}^2 + (v^I_F)^2 q_x^2} 
\eea
and
\bea
\chi(q_x,\omega \rightarrow 0) = \frac{K}{\pi v^I_F} \quad .
\label{charge_susc}
\eea
One can reach also at the same result through the thermodynamic Bethe ansatz 
equations at $T=0$. The susceptibility can be obtained as a second derivative of the ground state energy 
Eq.~(\ref{efundit}) with the external field. 
For related work see also Refs.~\onlinecite{haldane,giamarchi,aristov,konik}.

\subsubsection{Fermi velocity and Luttinger liquid parameter in gapped phase in the presence of chemical potential}

In the following we give the behavior of the Drude weight and charge susceptibility Eq.~(\ref{charge_susc}) in terms of 
$\beta^2$ and $H$ exploiting the Bethe ansatz equations for $v_F^I$ and $K$.
The Fermi velocity for solitons in the I phase (see Fig.~\ref{SG_LL_schemat}) can be found by taking the derivative of the 
soliton branch of the spectrum $\epsilon_s(P)$ with $P$, the momentum, 
\bea
v^I_F &=& \frac{\p \epsilon_s(P)}{\p P}\bigg|_{\epsilon_s=\epsilon_F} = \frac{\p \epsilon_s(\theta)}{\p \theta} 
\frac{1}{\p P /\p\theta} \bigg|_{\theta=B}  \quad ,
\\[3mm]
&=& \frac{\p \epsilon_s(\theta)}{\p \theta}\bigg|_{\theta=B} \frac{v_F}{2\pi \rho_s(B)}
\label{metallic_velocity}
\quad ,
\eea
where $v_F$ is the Fermi velocity of the commensurate phase and $\rho_s(B)$
is the soliton density at the Fermi point.

\begin{figure}
\unitlength=1mm
\begin{picture}(80,32)
\put(-3.5,5){\includegraphics[height=27mm]{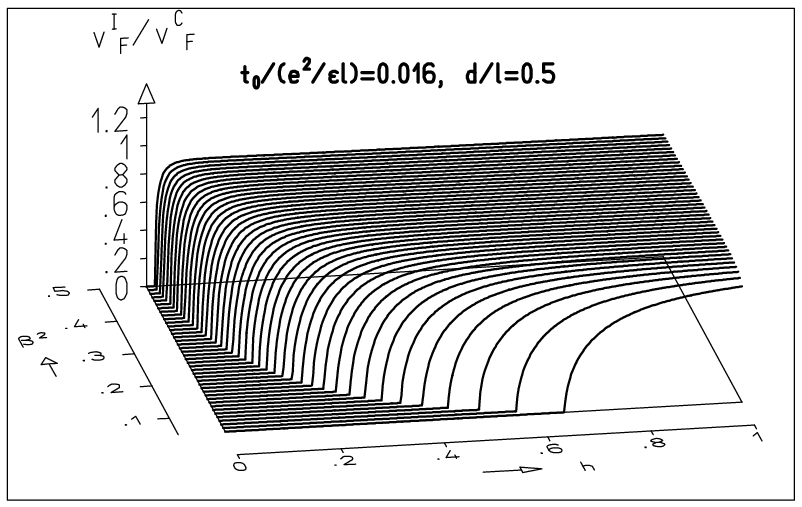}}
\put(39.95,5){\includegraphics[height=27mm]{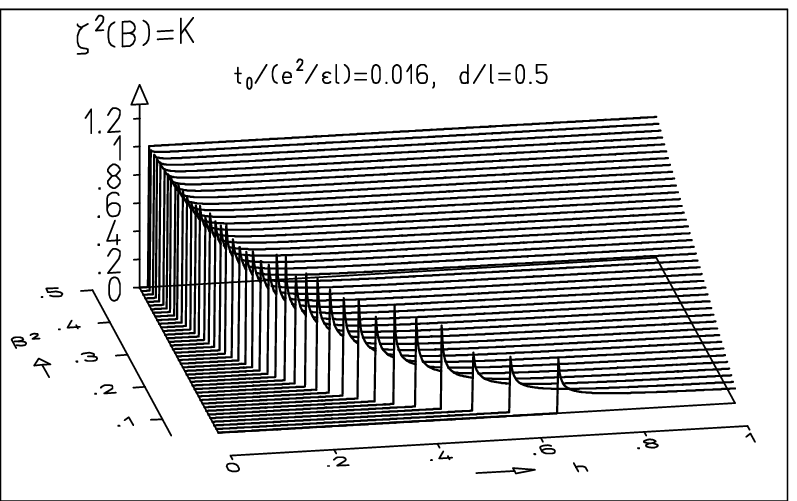}}
\end{picture}
\caption{
On the left panel the ratio of the Fermi velocity of I (metallic) phase as compared to that of the C (insulating) 
phase prior to the gap opening, $v_F^I/v_F=\p_\theta \epsilon_s(\theta)|_{\theta=B} /{2\pi \rho_s(B)}$ 
of Eq.~(\ref{metallic_velocity}), is shown. On right panel the dressed charge (LL parameter) as function of 
coupling constant, or equivalently, the interedge interaction strength $\beta^2$ and
external field is shown. The LL parameter is $K=1$ resembling free fermions at the phase transition point
when $H=M_s$ and takes the asymptotic value of $K=2\beta^2$ at larger values of $H$.
}
\label{velocity_and_charge}
\end{figure}

The equation for the energy spectrum is given in previous section by Eq.~(\ref{set-of-dress-e1}) 
and therefore $\p \epsilon_s/\p \theta$ is given by
\bea
\epsilon'_s(\theta)  + \int_{-B}^{B}\rd \theta' G(\theta-\theta') \epsilon_s'(\theta')
= M_s \sinh(\theta) \quad ,
\label{T_0_spectr}
\eea
whereas the particle density equation is the following
\bea
\rho_s(\theta) + \int_{-B}^{B}\rd \theta' G(\theta-\theta') \rho_s(\theta')
= M_s \cosh(\theta)/2\pi \; .
\label{density}
\eea
The Fermi points $\pm B$ are defined by $\epsilon(\pm B)=0$. We have solved these integral equations numerically and 
the graph for the Fermi velocity in the I (metallic) phase  as a function of the coupling constant and external field 
is given in Fig.~\ref{velocity_and_charge} (left)
%
whereas the Luttinger liquid parameter $K$ in the whole range of $H$ is shown in Fig.~\ref{velocity_and_charge} (right). 

From the above integral equations it is easy to obtain the asymptotic behaviors $H \rightarrow H_c$  
and $H/H_c\gg 1$, where $H_c =M_s$.
The behavior of the soliton Fermi velocity in the I phase is given by
\bea
\frac{v_F^I}{v_F} &=& \sqrt{2} \left( \frac{H}{H_c} - 1 \right)^{1/2}
\left[    
1-\frac{4\sqrt{2}G(0)}{3}\left(\frac{H}{H_c}-1 \right)^{1/2}
\right.
\nonumber
\\[3mm]
&& \left.
 - (2G^2(0)+1)\left(\frac{H}{H_c}-1 \right) +\ldots
\right]
\quad .
\eea 

Notice here that the square root behavior of the Fermi velocity $v_F^I$ is a generic feature
of all models with a gap in the spectrum whose low energy behavior fulfills Eq.~(\ref{quadratic_spectrum}). This leads 
to other generic properties in the behavior of the Drude weight and charge susceptibility below.

The parameter $K$ is given as the square of the dressed charge of the solitons $\zeta(\theta)$ around the Fermi points
$\theta=\pm B$,  which 
is defined \cite{korepin} as     
\bea
\zeta(\theta) =-\frac{\rd \epsilon_s(\theta)}{\rd H}  \quad ,
\eea
and as a result it is given by the integral equation
\bea
\zeta(\theta) +\int_{-B}^{B} \rd \theta' G(\theta-\theta')\zeta(\theta') = 1 \quad ,
\eea
$K=\zeta^2(B)$. This parameter characterizes the strength of the interactions of the excitations, and gives complete 
information on the exponents of various correlation functions in the metallic phase. The dependence of the LL parameter 
on the value of the external field is enforced by the field dependence of the Fermi boundary point $B(H)$.
 At the insulator-metallic phase transition point when $H=M_{\rm s}$ the integration interval
shrinks to a point and $K=\zeta^2(B\to 0)=1$.
The physical origin of this behavior is that at this point the density of solitonic particles of the theory 
is very low and they behave as free ones, leading thus to the characteristic $K=1$ value of the LL parameter.

At $H/M_s\rightarrow +\infty$ the Wiener-Hopf technique shows 
that the LL parameter $K\rightarrow 2\beta^2$. This coincides with $v_F^I \rightarrow v_F$.
Of course the limit of large values of the external field corresponds to the case of large density of solitons, and the system restores to properties 
prior to its gap opening. The $cos$ operator does not affect the properties of the system at these values of the external field.

\subsubsection{Asymptotic behavior of Drude weight and charge-charge correlations} 

I. ($H/H_c \to 1$).
Using the results of Ref.~\onlinecite{papa3} for the dependence of the LL parameter
$K$ on $H$ at $T=0$ for $H\rightarrow H_c$, the Drude weight behaves as
\bea
\frac{D}{v_F} &=& \sqrt{2} \left( \frac{H}{H_c} - 1 \right)^{1/2}
\left[
1-\frac{4\sqrt{2}G(0)}{3}\left(\frac{H}{H_c}-1 \right)^{1/2}
\right.
\nonumber
\\[3mm]
&& \left.
 - (2G^2(0)+1 +4\sqrt{2})\left(\frac{H}{H_c}-1 \right) +\ldots
\right]
\quad .
\eea
The Drude weight close to the critical point $H_c$ increases as a square root function of $(H/H_c-1)$. The same holds for the 
Fermi velocity in the I phase, $v_F^I$, close to $H_c$.

At $T=0$, the $H\rightarrow H_c$ asymptotic behavior for the charge susceptibility is
\bea
&&\chi(q_x,\omega \rightarrow 0) = 
\\[3mm]
&&\frac{1}{\sqrt{2}\pi v_F} \left(\frac{H}{H_c} - 1 \right)^{-1/2}
\left[ 1  
 + \frac{4\sqrt{2}G(0)}{3}\left(\frac{H}{H_c} - 1 \right)^{1/2} 
\right.
\nonumber
\\[3mm]
\nonumber
&&\left.
+ \left( 1 - 4 \sqrt{2}G(0) + \frac{50G^2(0)}{9} \right) \left(\frac{H}{H_c} - 1 \right) + \ldots 
\right] \,,
\eea
which clearly shows that it diverges as a square root function of $(H/H_c -1)$.
At small values of the coupling constant $\beta^2$ we expect the square root behavior for $D$ and $\chi$ to hold only for a 
very small interval of $H$ above $H_c$. 
For larger values of $H$ this behavior is expected to change to a logarithmic 
function of $(H/H_c -1)$.

II. ($H/M_s \gg 1$).
At $H/M_s \gg 1 $ it was shown previously in Ref.~\onlinecite{papa3} that the LL parameter 
$K\rightarrow 2\beta^2$. 
Therefore at $H\gg M_s$ we expect the experiments on the Drude weight and the charge susceptibility to give the following combinations of the coupling constant and Fermi velocity prior to gap opening
\bea
D = 2\beta^2 v_F \quad, \quad \chi=\frac{2\beta^2}{\pi v_F}\quad, \quad {\rm for} \quad H \gg M_s \quad .
\eea
(Here the usual $8\pi$ of Ref.~\onlinecite{papa3} is absorbed in $\beta^2$, which belongs now to $0 \le \beta^2 \le 1$).
Measuring therefore the Drude weight and the charge susceptibility at large values of the external potential should define the quantities needed for measuring the mass gap of the experiments of Kang {\it et al.} \cite{kang}.

\subsection{Optical conductivity in the insulating phase}

\begin{figure}
\unitlength=1mm
\begin{picture}(80,45)
\put(-70,65){\includegraphics[height=40mm]{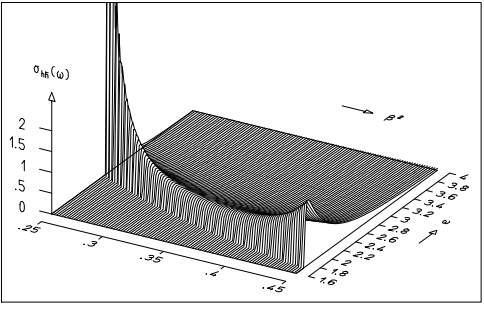}}
\end{picture}
\caption{The soliton-antisoliton contribution to the optical conductivity in
the C phase for the SG model.}
\label{sol-solconduc}
\end{figure}

 \begin{figure}
\unitlength=1mm
\begin{picture}(80,45)
\put(10,5){\includegraphics[height=40mm]{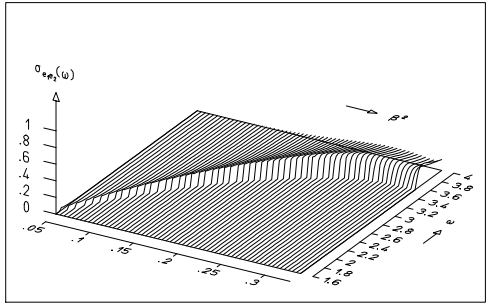}}
\end{picture}
\caption{The breather-breather ($e_1e_2$) contribution to the optical
conductivity in the C phase for the SG model.}
\label{b1-b2conduc}
\end{figure}

One important quantity to study experimentally in the context of the applicability of the sine-Gordon model in 
QHLJ systems is the optical conductivity. This quantity is quite useful since one can extract valuable 
information about the polarized excitations of the system. 

Just as in the metallic phase of the previous section, the optical conductivity in the insulating phase is defined as the Fourier transform 
of the current-current correlation function. 
Here one has to use the integrability property of the SG model, the form factor approach \cite{smirnov}.
The particles that contribute will be the ones that are odd under charge conjugation, the odd breathers $e_{2n-1}$ in the SG model.
The single breather contribution is simply given by a delta-peak at a
frequency given by the mass of the breather ($M_k = 2M_s {\rm sin}(\frac{\pi}{2}\xi k)$).
Other contributions will be the two-particle ones like, soliton-antisoliton and the odd-even breathers. At higher energies there will be 
contributions also form higher multiparticle states, like the 4-particle ones etc. 

The optical conductivity for the sine-Gordon model is calculated by Controzzi, Essler and Tsvelik \cite{essler}. We use here 
their results to illustrate  
plots of the first-second breathers and the soliton-antisoliton contributions to the optical conductivity 
as shown in Figs.~\ref{sol-solconduc} and \ref{b1-b2conduc}.
More detailed calculations for higher order terms are given in the appendix.

Let us take as an example the case of $\beta^2=1/4$, where there are two breathers in the spectrum.
As function of the frequency, the zero temperature optical conductivity will be zero for $\omega < M_1  = 2M_s \sin(\xi \pi/2 )$,
a delta-function peak at $\omega=M_1$ and then a continuous weight at $\omega> 2M_s$ for the soliton-antisoliton which comes before
the contribution from the first-second breather $\omega > M_1+M_2 = (1+\sqrt{3})M_s$.
 
At lower values of $\beta^2$ (stronger interedge forward interactions in the junction), the mass of the breathers decrease whereas 
the mass of the solitons increases, the number of breathers on the other hand increases and the odd-even breathers contributions 
will arise as continuum contributions until the soliton-antisoliton continuum contribution. 
Due to the heavier mass, the latter contribution in the optical conductivity will be much stronger than the one of the odd-even breathers. 
In the experiments, optical absorption at small frequencies will signify the presence of breathers in the spectrum and thus signify 
the presence of strong interedge Coulomb interactions in the line junction.

\subsection{QHLJ at filling fraction $\nu \sim 2$}

The QHLJ at filling fraction $\nu \sim 2$ in the case of the presence of a single impurity along the barrier 
has been discussed beautifully  by Kim and Fradkin in Ref.~\onlinecite{kim}. Here we are mainly interested 
in the case when interedge tunneling can take place at any point along the barrier but largely follow the ideas of Ref.~\onlinecite{kim}.

The calculations of the previous sections can be extended also to the spin-full case, when 
the 2DEG subsystems on the sides of the barrier are at filling fractions $\nu \sim 2$.
The Hamiltonian in these 1D systems separates into two commuting
parts, the spin and charge ones ${\cal H}= {\cal H}_{\rm c} + {\cal H}_{\rm s}$. 
We focus here on the case of a gapful spin sector, where the chemical potential for the topological 
charges consists naturally on the Zeeman coupling term.
Information on the strength of electron interactions can also be obtained by working in this sector. 
We discuss here the possibility for observation of the SILL phase that has been found to 
exist in 1D systems with spin-charge energy scale separation \cite{fiete}.

The edge states of a 2DEG on a QH plateau at bulk filling $\nu \sim 2$ 
are composed of spin up and spin down branches. To account for both of them 
we need to introduce two bosonic fields, 
$\phi_\sigma$, where $\sigma=\uparrow,\downarrow$ are the up and down spin projections. 
The right and the left movers are represented by $R_\sigma\sim e^{-i\sqrt{4\pi}\phi_\sigma}$, 
and $L_\sigma\sim e^{i\sqrt{4\pi}\bar{\phi}_\sigma}$, 
respectively. The electron spin brings about changes in the dynamics --- terms like the spin-spin 
exchange interactions as well as the Zeeman interactions have to be added into the Hamiltonian. 
Introducing the scalar U(1) and SU(2) vector currents $J = :R_\sigma^\dagger R_\sigma:$, 
${\bf J}= :R^\dagger_\alpha (\vec \sigma_{\alpha\beta}/2)R_\beta:$, and similarly for the left movers, $R\to L$,
the charge and spin components of the Hamiltonian (without the Zeeman term)
can be written as
\bea
{\cal H}_c=\frac{\pi v_c}{2} (JJ+\bar J \bar J + g_c J \bar J) \quad ,
\\[2mm]
{\cal H}_s =\frac{2\pi v_s}{3} \left({\bf J}{\bf J} + {\bf \bar J}{\bf \bar J} + 
g_s {\bf J}{\bf \bar J} \right) \quad .
\eea

In the standard way (Ref.~\onlinecite{tsvelik}) we introduce the charge and the spin bosonic fields
$\phi_i=(\phi_\uparrow \pm \phi_\downarrow)/\sqrt{2}$
for $i=c,s$ and similarly for the opposite chirality fields
$\bar{\phi}_i=(\bar{\phi}_\uparrow \pm \bar{\phi}_\downarrow)/\sqrt{2}$.
The charge excitations remain free and described by the LL model.
\begin{equation}
H_c =  \frac{\hbar v_c}{2}\int \rd x \left[K_c(\p_x \Phi_c )^2 + \frac{1}{K_c}(\p_x \Theta_c)^2\right]
\quad,
\label{H_charge}
\end{equation}
where $\Phi_c = \phi_c +\bar{\phi}_c$ and $\Theta_c=\phi_c-\bar{\phi}_c$.
The Fermi velocity is renormalized by the $g_c$ density-density type interactions. 
(The relationship between $v_c$, $K_c$ and $g_c$ is nonuniversal --- when the
interactions are weak these parameters are related by simple expressions (Ref.~\onlinecite{tsvelik})].

In $H_s$ the $g_s$ type perturbation is subject to renormalization group (RG) flow. ${\bf J}{\bf \bar J}$ is a marginal 
operator and the sign of interactions is important in that 
it can flow the system toward the strong coupling phase, with no scale invariance, or it can scale it to zero
(the gapless phase).
In general the spin-spin exchange interaction $g_s {\bf J}{\bf \bar J}$ can be extended to an 
anisotropic one
\bea
g_s{\bf J}{\bf \bar J} \to g_\parallel J^z \bar J^z +\frac{1}{2}g_\perp\left(J^+\bar J^- + h.c.\right)
\quad .
\eea
This can be the case in the samples of the experiments of {\em Kang et al.} \cite{kang}.
The new constants are: $g_\perp=g_s$ the $XY$ component of exchange interaction, and $g_\parallel=(1+\lambda)g_s$ the Ising type coupling. $\lambda$ is the
strength of the Ising anisotropy. 
The magnetic field introduces a Zeeman type coupling [Eq.~\ref{Zeeman}].
The spin part of the Hamiltonian is described by the sine-Gordon model, 
where the $cos$ operator is introduced by the $XY$-type spin-flip backscattering 
interaction $g_\perp$. The RG flow of the model is discussed in detail in Ref.~\onlinecite{kim}.
The isotropic case ($\lambda=0$) is subject to SU(2) symmetric RG flow, with perturbation marginally irrelevant at the ferromagnetic case 
$g_s<0$, and marginally relevant in the antiferromagnetic 
case $g_s > 0$. The latter type of interaction generates 
a gap in the spin sector.
A gap in the spin sector can open up even in the case of ferromagnetic 
type interactions, when the magnetic interactions are anisotropic. 
This happens when $\lambda < 0$.  Either of these cases would be of interest 
in the context of our previous discussion.

The Zeeman coupling (magnetic field) can be regarded as a field that is coupled with the 
topological charges of the SG model 
\bea
H_s  &=&  \frac{\hbar v_s}{2}  \int \rd x \left[K_s(\p_x \Phi_s )^2 + \frac{1}{K_s} (\p_x \Theta_s)^2\right]
\\[3mm]\nonumber
  && \hspace{35mm}+ 2\mu_s
\cos(\sqrt{8\pi }\Phi_s)
\,,
\label{H_spin}
\eea

We discuss here various cases that arise concerning electron tunneling conductance across the
junction. 

In systems where the electron interactions are very strong, 
a separation in the energy scales in spin and charge sectors arises. 
At low temperatures it is possible for the
energy to be low compared with the characteristic charge energy and high compared with the
characteristic spin energy, $E_s \ll k_B T \ll E_c$. 
In this window of energy, the charge sector is very close to the ground state whereas the spin sector is thermally excited.
As it was shown in Ref.~\onlinecite{fiete} this state has quite interesting properties, particularly the correlation functions decay 
exponentially in space and the exponents of the tunneling density of states is also different from a simple Luttinger liquid.
In the QHLJ systems that we examine here, such conditions can arise  naturally. 
Due to the small barrier widths, less than a magnetic
length $l_B$ in the systems of Ref.~\onlinecite{kang},
we believe that Coulomb interactions are strong whereas the energy scale in the
spin sector can be varied by tuning the magnetic field.
The charge energy scale is of the order of $E_c \sim e^2/\epsilon l_B \approx 10.76{\rm meV}$. On the other hand, the
spin excitations in the case of the antiferromagnetic $g_s > 0$ interactions are gapped. At values of the
Zeeman term bigger than a critical value $H> H_s$, the gap closes. It is at values of the Zeeman field
only slightly in excess of $Q_c$ 
that the SILL phase can occur.
At $H \to H_s$ from above, the characteristic Fermi velocity of excitations in spin sector
becomes very small, $v_s \sim v_F[(H/H_s-1)^{1/2} +\ldots] \to 0$ and the energy scale in spin sector vanishes $E_s \sim \hbar v_s/l_B \to 0$. 

As discussed in Ref.~\onlinecite{kim}, the most relevant tunneling operators that 
account for both charge and spin transport across the junction are 
the electron tunneling
$\hat{O}_{\rm el}= R_{\uparrow}^\dagger L_{\uparrow} + R_{\downarrow}^\dagger L_{\downarrow} + {\rm h.c.} \sim 
\cos(\sqrt{2\pi K_c} \Phi_c)\cos(\sqrt{2\pi K_s} \Phi_s)$, the singlet pair tunneling
$\hat{O}_{\rm pair} = R_{\uparrow}^\dagger R_\downarrow^\dagger L_\downarrow L_\uparrow + {\rm h.c.} 
\sim\cos(\sqrt{8\pi K_c}\Phi_c)$, and the triplet pair tunneling
$\hat{O}_{\rm s} = R_\uparrow^\dagger L_\uparrow L_\downarrow^\dagger R_\downarrow + {\rm h.c.}\sim 
\cos (\sqrt{8\pi K_s}\Phi_s)$.

We analyze in the following the relevance of the tunneling operators at various values of the Zeeman term 
in the spin gapped state.
Since the spatial separation of the counterpropagating edges on the sides of the barrier is of the order of one magnetic length 
or less we assume interedge tunneling can take place at any point along the barrier. The scaling dimensions of the operators 
depend on the LL parameters of the charge and spin sectors and are given by \cite{kim} $d_{\rm el} = (K_c + K_s)/2$, whereas 
$d_{\rm pair} = 2 K_c$ and $d_{\rm s} = 2 K_s$. 
First of all, in the spin gapped phase the only allowed tunneling process is the singlet pair tunneling $\hat{O}_{\rm pair}$ which
is independent of the spin bosonic field. The spin gap is generated by the operator $\cos(\sqrt{8\pi K_s} \Phi_s)$ [either when $g_s<0$ and any $\lambda$, 
or when $g_s>0$ and $\lambda<0$], and therefore $\Phi_s$
takes as value one of the minima of the potential. The average of the $\cos(\sqrt{2\pi K_s}\Phi_s)$ will alternate in this set and result in 
a vanishing average. $\hat{O}_{\rm pair}$ term will lead to a gap opening in the charge sector whenever $K_c < 1$. The charge and spin 
currents will be transmitted across the barrier.
  By sweeping the Zeeman term to higher values than the spin gap in the 
region $H\sim H_s$, at finite temperatures, due to vanishing energy scale in spin sector, the SILL should arise. Due to thermal excitation 
the spin correlation functions here again decay exponentially and spin tunneling 
processes are suppressed. The only allowed tunneling processes will be again the singlet pair tunneling although this will be 
relevant only at $K_c < 1/2$. The jump in scaling dimensions of this gap generating operator in charge sector 
is a signature of the SILL state.
A perfect charge transport at these values of the Zeeman term will signify strong correlations along the barrier. At higher
values of the magnetic field  the pair tunneling operator and the electron tunneling operator will be both relevant. At strong interactions
we expect $\hat{O}_{\rm pair}$ to be more relevant then $\hat{O}_{\rm el}$. Charge and spin transport will display a peak as function 
of the bias voltage.

In the presence of a single impurity the phase diagram as function of $K_c$, 
$K_s$, and magnetic anisotropy, was studied in Ref.~\onlinecite{kim}.
Here we add that in the spin gapped phase in the presence of a single impurity and the SILL phase ($H\sim H_s$)
a divergence on the density of states at the impurity 
point arises whenever $K_c > 1/4$ and charge tunneling through the barrier at low bias voltages 
does not occur.
In the opposite case at $K_c <1/4$ the presence of the impurity will lead to a suppression of the density of states 
and a peak in the charge conductance through the barrier should be observed.

\section{Conclusions}

In this paper we studied the properties of QHLJ created by the narrow cleaved edge overgrowth barriers.
Such line junction have a sort of properties that make them be advantageous to other sorts of line 
junctions, as for instance, the gate created junctions, 
in that, first, the barriers present to the electrons sufficiently abrupt potentials 
that edge reconstruction phenomena do not occur. 
This allows for the edge states of the CEO created line junctions to be described by single chiral channels.
Second, these barriers can be narrow enough to allow for electron tunneling to take place at any 
point along the length of the barrier, leading to low energy description in terms of generalized 
sine-Gordon models. And third, these systems allow for strong Coulomb interactions that 
lead to pronounced features in transport experiments. 

We discussed here ways of estimating the parameters of the field theory model that describes the QHLJ 
system\cite{kang} at low energies. These parameters are hard to estimate with accuracy 
from microscopic models.
Therefore we undertook an indirect way for their estimation. We showed that information on 
the value of the LL parameter and the Fermi velocity can be extracted by using the integrability property
of the sine-Gordon model. When forward scattering repulsions are present the SG model is characterized 
by a gap in the spectrum. The excitations consist on topological particles and their bound states 
(if Coulomb interactions are strong). The gap 
can close in the presence of fields that couple with these charges. 
In the so obtained metallic phase we propose the study of the charge susceptibility and the Drude weight.
It turns out (see Figs.~\ref{modes}, \ref{velocity_and_charge}) 
that not very far from the critical value of the gap closing, the LL parameter which depends on 
$H$ and $\beta$, reaches as its asymptotic value, the one that would characterize the system prior to gap 
opening. In the insulating phase, to obtain information on the properties of the system, we propose the 
measurement of the optical conductivity of the line junction.
The optical conductivity provides valuable information on the polarized excitation of the system.
Delta-function type features appear in optical conductivity  for each particle type 
odd under charge conjugation (all odd breathers $e_{2n-1}$),
 as well as two and more particle bands appear in the spectrum.
This measurement should therefore lead to information about the number of particles present, and 
thus the strength of the interactions.
Here the form factor formalism\cite{smirnov} is used. 

We discussed here also the case when spin degrees of freedom take part in the dynamics, i.e. the case of 
bulk-filling fraction $\nu \sim 2$ on the 2DEG subsystems. Most interesting is the study of the case a spin 
gapped phase. In this case the Zeeman term can be tuned to values just above the mass gap, resulting in
a strong separation of charge and spin energy scales. 
At finite temperatures, much less than the charge energy scale but larger than the spin energy scale, 
the SILL state can be realized. Such a state has been proposed to exist\cite{fiete} in 1D systems with 
strong Coulomb interactions -- Wigner crystals in charge sector and weak spin exchange interactions, 
but its observation is yet to be seen. Its unusual properties stem from the reduction of the spin energy scale,
which turns out to have important effects on the correlation functions and thus transport properties 
of this state.

We focused here on the coherent line junctions where tunneling takes place at all points along the junction.
At wider barriers, due to momentum conservation, interedge tunneling is prohibited except at points of impurities along the barrier. 
The low energy physics of systems with just a single impurity can be mapped to the boundary sine-Gordon model (Ref.~\onlinecite{kim}), 
which allows in principle for an exact calculation of the differential conductance. In real experiments however, a temperature and 
voltage dependence of the tunneling amplitude occurs. The tunneling conductance calculated with the Bethe ansatz taking into account   
these other effects will be discussed in a future publication \cite{papa_kim}.

\section*{ACKNOWLEDGEMENTS}

The author is grateful to Dave Allen, Matthew Grayson, Woowon Kang, Allan MacDonald, Stefano Roddaro, 
Tilo Stroh, and Alexei Tsvelik for illuminating conversations. The author is especially 
grateful to Allan MacDonald for teaching him the Hartree-Fock theory in the context of line junctions 
in QH systems and many other illuminating discussions and a careful reading of the manuscript and 
to Eun-Ah Kim for valuable comments on the manuscript and discussions. 
This work was supported by the National Science Foundation under Grant No. DMR-0412956.

\section{Appendix}

\subsection{TBA 
equations for purely elastic theories and the $T=0$ limit for the sine-Gordon model}

Here we write the thermodynamic Bethe-Ansatz (TBA) equations for systems for which there is no reflection on 
scattering of the particles of their  spectrum, i.e. we consider only diagonal scattering or purely elastic theories. 
These equations apply for the SG model only at particular values of the coupling constant, namely 
$\beta^2 = 1/n$, where $n$ is an integer or, in the presence of the external fields, only at temperatures much 
lower than the size of the gap, $T \ll M_s$. At zero temperature, which is the case of our interest in (\ref{drude}),  
the BA equations for SG in the presence of external fields can be derived only based on the fact that in case 
$H \ge h_c =M_s$ in the ground state there will appear a soliton condensate and therefore 
the equations with a single kind of particles in the spectrum apply.
The $T=0$ will be good for every value of the coupling constant of SG.

We are dealing here with a Lorentz invariant model and in this case it is usual to use the 
following parametrization for the particles's energy and momenta
\bea
E = m \cosh\theta \quad , \quad P = (m/v_F) \sinh\theta \quad ,
\eea
where $m$ is the mass of the particle and $v_F$ the Fermi velocity [corresponding to the Fermi velocity of the C 
phase of Eq.~(\ref{H_short_ranged})].
In this parametrization Lorentz invariant quantities like the scattering matrix elements will depend only on 
the differences of rapidities (invariant under Lorentz boosts which shift the rapidities by the
same constant).
Suppose we have N particles of mass $m$ on a closed circle of length $L$.
Let the two-body scattering matrix of particles with momenta $p_1$ and $p_2$ be equal to
$S(\theta_1-\theta_2)=\exp\{i\varphi(\theta_1 - \theta_2)\}$. If we consider a particle of momentum $p_j$
traveling around the circle, on returning to the initial point its wave function acquires the factor 
\bea
\exp\Bigl[ip_i L + i \sum_{j\neq i} \varphi(p_i,p_j)\Bigr] \quad,
\eea
and this has to be equal to unity. Therefore the argument of the exponential has to fulfill
\bea
e^{ip_i L} \prod_{j\neq i} S(\theta_i - \theta_j) = 1 \quad,
\eea
or
\bea
mL\sinh\theta_i +\sum_{j\neq i} \phi(\theta_i - \theta_j) = 2\pi n_i \quad .
\label{periodic_BC}
\eea
There are $N$ different integer numbers $n_i$ (eigennumbers) which determine the eigenvalues of the Hamiltonian
\bea
{\cal E}=\sum_{i} m \cosh\theta_i \quad,
\eea
In the thermodynamic limit $L\rightarrow \infty$, the distance between the rapidities, solutions of Eq.~(\ref{periodic_BC}), 
is of the order of $1/L$ and in this case one introduces the densities of rapidities occupied by particles 
$\rho_p(\theta)$ and the density of holes $\rho_h(\theta)$ (rapidities which can potentially be occupied by 
particles).
Taking the derivative of (\ref{periodic_BC}) with $\theta_i$ 
one gets the following integral equations for the dressed distributions
\be
\rho_p(\theta) + \rho_h(\theta)=
\frac{m \cosh \theta}{2\pi}+\left(K *
\rho_p\right)(\theta) \quad ,
\label{distrib}
\ee
and for the more general case of several types of particles $a$ with $a=1,2, \ldots,n$, with masses $m_a$
\be
\rho_p^a(\theta) + \rho_h^a(\theta)=
\frac{m_a \cosh \theta}{2\pi}+\sum_{b=1}^n \left(K_{ab} *
\rho_p^b\right)(\theta) \quad ,
\label{distrib}
\ee
where $(m_a \cosh \theta)/2\pi =(1/2\pi)d p_a^0/d \theta \equiv \rho_{a,p}^0(\theta)$
is the bare distribution function of the particles of type `$a$',
whereas the Kernel $K$ is related to the two-body scattering matrix
\be
K_{ab}(\theta)= \frac{1}{2\pi i} \frac{d}{d \theta} \ln S_{ab}(\theta)
\quad .
\ee
The energy of the system is given by
\be
\label{E}
{\cal E}= \int \sum_a \rho_p^a(\theta) m_a \cosh \theta \, \rd \theta
\quad ,
\ee
and in the presence of a chemical potential coupled to the ``charge'' of
the particles it is
\be
\bar{\cal E}= {\cal E}-\sum_a h_a q_a \quad .
\ee
In Eq.~(\ref{E}) $m_a \cosh \theta = \epsilon_0(\theta)$, the bare
energy.

The entropy of the distribution between $\theta$ and $\theta +d
\theta$ is the logarithm of the number of distributions
\be
d S = \ln \frac{\left\{L[\rho_p^a(\theta)+\rho_h^a(\theta)] d \theta\right\}!}
{[L\rho_p^a(\theta)]! [L\rho_h^a(\theta)]!} \quad.
\ee
Here one uses the Stirling formula $\ln(n!) \simeq n(\ln n-1)$. Then
the entropy per unit length is
\bea
S/L=&& \int \sum_a \left\{
 [\rho_p^a(\theta)+\rho_h^a(\theta)] \ln
\left[ \rho_p^a(\theta)+\rho_h^a(\theta) \right]
\right.
\nonumber\\[3mm]
&&\left.
 -\rho_p^a(\theta)\ln\rho_p^a(\theta) - \rho_h^a(\theta)\ln\rho_h^a(\theta)
\right\} \rd \theta
\quad .
\eea
From the condition of the minimum of the free energy $F=E-TS$
in the equilibrium
state one gets relations between $\delta \rho_p^a$ and $\delta
\rho_h^a$, which when substituted in (\ref{distrib}) lead to the following set
of integral equations
\be
m_a \cosh \theta -h_a q_a = \epsilon_a(\theta) +T\sum_{b=1}^n
K_{ab}*\ln \left( 1+e^{-\epsilon_b/T} \right) \quad ,
\label{set-ofequat}
\ee
where the functions $\epsilon_a(\theta)$ are defined by
\be
\frac{\rho_p^a(\theta)}{\rho_h^a(\theta)} =
\exp\left\{-\frac{\epsilon_a(\theta)}{T}\right\} \quad .
\ee
Using the dressed energies one finds a formula for the free
energy per unit length
\be
F/L= - T \sum_a m_a \int \frac{\rd \theta}{2\pi}\cosh \theta
\ln\left[1+e^{-\epsilon_a(\theta)/T}\right] \; .
\label{free-energy}
\ee
As $T\rightarrow 0$ the equations for the dressed energies (\ref{set-ofequat})
reduce to
\be
m_a \cosh \theta -h_a q_a = \epsilon_a(\theta) - \sum_{b=1}^n
K_{ab}*\epsilon_b^{-} \quad ,
\label{set-of-dress-e}
\ee
where $\epsilon_a^{-}(\theta)$ is the negative part of the dressed energy
$\epsilon_a(\theta)$
[usually between the interval we note $(-B,B)$].
If from the set of the particles only one has positive charge, then
all the energies except $\epsilon_1(\theta)$ are positive everywhere
and the set of the equations (\ref{set-of-dress-e}) reduces further to
\be
\epsilon_1(\theta) - \int_{-B}^{B} \rd \theta' K_{11}(\theta-\theta')
\epsilon_1(\theta') = m_1\cosh \theta - h_1 q_1 \quad ,
\label{spectrum_T_0}
\ee
where $\epsilon_1(\pm B)=0$.
From Eq.~(\ref{free-energy}) one finds the soliton (particle $a=1$)
contribution to the ground state energy of the system
\be
\label{efundit}
{\cal E}_0
= \int_{-B}^{B} \rd \theta \, \rho_{1,p}^0(\theta) \,
\epsilon_1(\theta)
= m_1\int_{-B}^{B} \frac{\rd \theta}{2\pi} \cosh \theta  \,
\epsilon_1(\theta) \quad .
\ee
In Eq.~(\ref{efundit}), $\rho_{1,p}^0(\theta)$ is the bare distribution of
particles of type `$1$' in the ground state.

In the following we give the 2-particle soliton-soliton and soliton-antisoliton scattering 
matrices for the SG model\cite{zam_77,zam_79} 
\bea
&& S_{ss}^{ss}(\theta) = S_{\bar{s}\bar{s}}^{\bar{s}\bar{s}}(\theta)
=S_0(\theta) \quad ,
\\[3mm]\nonumber
&& S_{s\bar{s}}^{s\bar{s}}(\theta) = S_{\bar{s}s}^{\bar{s}s}(\theta)
=-\frac{ \sinh (\theta/ \xi) }{ \sinh \frac{1}{\xi}(\theta - i\pi)} S_0(\theta) \quad {\rm Trans.}
\\[3mm]\nonumber
&& S_{\bar{s}s}^{s\bar{s}}(\theta) = S_{s\bar{s}}^{\bar{s}s}(\theta)
=-\frac{ \sinh (i\pi/ \xi) }{ \sinh \frac{1}{\xi}(\theta - i\pi)} S_0(\theta) \quad {\rm Refl.}
\eea
with
\bea
\nonumber
S_0(\theta) =-\exp\left[-i \int_0^\infty \frac{\sin(\theta t /\pi) \sinh(\frac{1 -\xi}{2}t)}
{t\cosh(t/2) \sinh(\xi t /2)} \rd t
\right] \quad .
\eea
At $\xi=1/n$ the $S_{\bar{s}s}^{s\bar{s}}(\theta)=S_{s\bar{s}}^{\bar{s}s}(\theta)=0$ and the scattering is 
purely transmissive. In the above equations
\bea
\xi = \frac{\beta^2}{1-\beta^2} \quad .
\eea
At $\beta^2=1/2$, which corresponds to the fermion free point, $S_{ss}=-1$ and $S_{s \bar{s}}=S_0=-1$.

\subsection{Kernels of particle interactions at $\beta^2=1/3$ and $\beta^2=1/5$}

To calculate the interacting kernels of the particles we use the following formulas 
(Ref.~\onlinecite{gradshteyn}) 
\begin{equation}
\frac{\sinh(n\xi)}{\sinh\xi} = \sum_{k=0}^{[(n-1)/2]} (-1)^k \textbinm{n-k-1}{k} 2^{n-2k-1}\cosh^{n-2k -1}\xi
\;,
\end{equation}
and  (Ref.~\onlinecite{gradshteyn}) 
\bea
\int_0^{\infty} \rd \xi \cos a \xi \frac{\cosh \beta \xi}{\cosh \gamma \xi} = \frac{\pi}{\gamma} 
\frac{ \cos \frac{\beta \pi}{2\gamma} \cosh \frac{a\pi}{2\gamma}}{
\cosh \frac{a\pi}{\gamma} + \cos \frac{\beta \pi}{\gamma}}
\quad ,
\eea
for [$|{\rm Re}\beta|< {\rm Re}\gamma$, for real $a$].

For the $\beta^2=1/3$, the interparticle interacting kernels are
\bea
&& K_{bs}(\theta) = 
-\frac{\sqrt{2}}{\pi}\frac{\cosh\theta}{\cosh(2\theta)} \quad ,
\\[3mm]
K_{ss}(\theta)  
&=&-\frac{1}{2\pi}\frac{1}{\cosh\theta} \; ,\; 
K_{bb}(\theta)= -\frac{1}{\pi}\frac{1}{\cosh\theta} \; .
\eea

For the $\beta^2=1/5$, the interparticle interacting kernels are

\bea
K_{ss}(\theta) &=& -\frac{1}{\pi} \left(\frac{1}{\cosh\theta} + \frac{2\sqrt{2}\cosh\theta}{\cosh(2\theta)} \right)
\quad ,
\\[3mm]
K_{1s}(\theta) &=& -\frac{1}{\pi} \left(\frac{4\cos(\pi/8)\cosh\theta}{\cos(\pi/4)+\cosh(2\theta)} \right)
\quad ,
\\[3mm]
K_{2s}(\theta) &=& -\frac{1}{\pi} \left(\frac{2}{\cosh\theta} + \frac{2\sqrt{2}\cosh\theta}{\cosh(2\theta)} \right)
\quad ,
\\[3mm]
K_{3s}(\theta) &=& \frac{1}{2} \left( K_{1s} + K_{12} \right)
\quad ,
\\[3mm]
K_{11}(\theta) &=& -\frac{1}{\pi} \left( \frac{2\sqrt{2}\cosh\theta}{\cosh(2\theta)} \right)
\quad ,
\\[3mm]
K_{22}(\theta) &=& -\frac{2}{\pi} \left(\frac{1}{\cosh\theta} + \frac{2\sqrt{2}\cosh\theta}{\cosh(2\theta)} \right)
\quad ,
\\[3mm]
K_{33}(\theta) &=& -\frac{1}{\pi} \left(\frac{4}{\cosh\theta} + \frac{6\sqrt{2}\cosh\theta}{\cosh(2\theta)} \right)
\quad .
\\[3mm] \nonumber
\eea

\subsection{Optical conductivity of SG model in the gapful regime}

In the massless regime  we computed 
physical quantities like the Drude weight and the optical conductivity.
Similar calculations can also be done in the massive regime. 
Here the integrability property of the sine-Gordon model can be used. Indeed, 
in addition to the energy spectrum that is obtained through the Bethe ansatz, 
quantities like the form factors and correlation functions can be
computed exactly based on the bootstrap formalism. In this section, the low
frequency optical conductivity is computed exactly using the form factors.
Similar calculation has been done previously\cite{essler} 
to determine the optical conductivity of the Mott-Hubbard insulator. 
We try in this appendix to extend their work to compute additional
breather-breather contribution.

The optical conductivity is defined as the Fourier transform of the
current-current correlation function. It is an important physical quantity
that provides information on the polarized excitations of a system.
Explicitly, the relationship between the optical conductivity and the
current-current correlation function is given by:
\begin{eqnarray}
\sigma(\omega,q) &=& \frac{1}{\omega}{\rm Im}\Bigl(
i\int_{-\infty}^{+\infty}dx \int_0^\infty dt e^{i(\omega
+i\epsilon)-iqx}\nonumber\\ [3mm]
&& \times \Bigl. [j(x,t),j(0,0)]\Bigr) \quad .
\end{eqnarray}
The current-current correlation function, can be computed by
introducing the identity resolution, as follows
\begin{eqnarray}
\langle j(x,t)j(0,0)\rangle &=& \sum_{n=0}^\infty\sum_{a_i}\int
{d\theta_1 \ldots d\theta_n \over (2\pi)^n n!}e^{i\sum_{j=1}^n
(p_jx-\epsilon_jt)}
\nonumber\\ [2mm]
&& \left| \langle 0|j(0,0)|\theta_1\ldots\theta_n\rangle_{a_1\ldots a_n}\right|^2 \quad ,
\end{eqnarray}
 where $|\theta_1\ldots\theta_n\rangle_{a_1\ldots a_n}$ is a $n$ particles
state with rapidities $\theta_1\ldots\theta_n$. The indices $a_i$ take for
value $\pm 1/2$ corresponding to a soliton or anti-soliton or $a_i=1,\ldots,
[1/\xi]$ corresponding to the different breathers.  $p_j$ and $\epsilon_j$
are, respectively the momentum and energy of the $j$th particle. The link
between these variables and the rapidity is given by Eq.~(42). The matrix
element $\langle 0|j(0,0)|\theta_1\ldots\theta_n\rangle_{a_1\ldots a_n}$ is
called {\it form factor} and will be noted $f_{a_1\cdots
a_n}(\theta_1\ldots\theta_n)$. The knowledge of all form factors leads to the
current-current correlation function and thus to the optical conductivity.
From the previous definitions, one easily deduces the following relation:
\bea
&&\sigma(\omega,q) = \frac{1}{\omega} \Im{\rm m} \left(-2\pi
\sum_{n=0}^{\infty}\sum_{a_i} \int \frac{\rd \theta_1 \ldots \rd
\theta_n}{(2\pi)^n n!}
\right.\nonumber \\[3mm]
&&\left.
\Bigl| f_{a_1\cdots a_n}(\theta_1 \ldots \theta_n)\Bigr|^2
\left[\frac{\delta(q-\sum_j M_j \sinh\theta_j/v_F)} {\omega
+i\epsilon-\sum_j M_j \cosh\theta_j} -
\right.\right.
\nonumber\\[3mm]
&&\left.\left.
\hspace{20mm}
-\frac{\delta(q+\sum_j M_j \sinh\theta_j/v_F)}{\omega +i\epsilon+\sum_j M_j
\cosh\theta_j}
 \frac{}{}\right]\right) \; .
\eea
In the limit $\epsilon\rightarrow 0$, the imaginary part can be computed
leading to an additional Dirac function over the frequency. The one-particle
and two-particle states contribution to the conductivity at vanishing
momentum is given by:
\begin{eqnarray}
&&\sigma(\omega,q=0) \approx  \sum_{a}{\pi v_{F}^{C} \over M^2_{a}}
|f_a(\theta=0)|^2 \delta(\omega-M_a)\\\nonumber
&& +\sum_{a_1,a_2} {2\pi v_F^C \over \omega}
{|f_{a_1,a_2}(\theta_1,\theta_2)|^2\Theta(\omega-M_1-M_2)\over \left[
(\omega^2-(M_1-M_2)^2)(\omega^2-(M_1+M_2)^2)\right]} \, .
\label{} 
\end{eqnarray}
The particle rapidities in the two particle contributions are related to
each other and to the frequency by the following relations:
\begin{eqnarray}
M_1 {\rm sinh}\theta_1 + M_2{\rm sinh}\theta_2 &=& 0  \quad , \\ [3mm]
M_1 {\rm cosh}\theta_1 + M_2{\rm cosh}\theta_2 &=& \omega \quad .
\eea
The limitation of the computation to the one particle and two-particles states
is valid at least at low frequency, $\omega$. More explicitly, here are some
rules to determine the form factors required for the computation of the
conductivity:
\begin{itemize}
\item Only the form factors, $\langle
0|j(0,0)|\theta_1\ldots\theta_n\rangle_{a_1\ldots a_n}$, where the sum of
the $n$ particles mass is less than $\omega$ will contribute to
$\sigma(\omega,q)$.
\item The current, $j(x,t)$, being odd under charge conjugation while the
ground state, $|0\rangle$ is even, only the states,  $\langle 0| j(0,0)
|\theta_1\ldots\theta_n\rangle_{a_1\cdots a_n}$, odd under charge
conjugation will lead to a non-vanishing form factor.
\end{itemize}
The one and two-particle states odd under charge conjugation are the following:
\begin{itemize}
\item $|\theta_1,\theta_2\rangle_{1/2,-1/2} -
|\theta_1,\theta_2\rangle_{-1/2,1/2}$
\item $|\theta\rangle_n$ where $n$ is odd.
\item $|\theta_1,\theta_2\rangle_{r,s}$ where $r+s$ is odd.
\end{itemize}
It is not surprising that the lowest breather ($n=1$) is odd under charge
conjugation. The soliton and anti-soliton being fermions, the multi-solitons
states have to be odd under particle exchange. This holds also for bound
states like the breather. It implies that if the state is odd under charge
conjugation, it has to be even under parity transformation and vice versa.
However, since there is an attractive short range interaction between the
particles, the lowest bound state (or breather) is obtained when the
particles position overlaps the most. This overlap is maximized for the
states that are even under parity and thus odd under charge conjugation.
The soliton-antisoliton form factor has been known for a long time.
Explicitly, it is given by:
\begin{equation}
\langle 0| j(0,0) |\theta_1,\theta_2\rangle_{s\overline{s}} =
\frac{M_s}{\pi}{{\rm cosh}\frac{\theta_{+}}{2}\over {\rm
cosh}\frac{\theta_{-}+i\pi}{2\xi}}F(\theta_{-})\; ,
\label{soliton-solitonFF}\end{equation}
where $M_s$ is the soliton mass, $\theta_{\pm} = \theta_1\pm\theta_2$, and $F$
is called the minimal form factor which is given by:
\begin{eqnarray}
F(\theta) \hspace{-1.5mm} &=& \hspace{-1.5mm} {\rm sinh}\left(\frac{\theta}{2}\right) 
\\ [3mm] \nonumber
&\times& \hspace{-2mm}{\rm exp}\left[ -\int_0^\infty {{\rm sinh}^2(t(1-i\theta/\pi)){\rm
sinh}((1-\xi) t)\over {\rm sinh}(\xi t){\rm sinh}(2t){\rm
cosh}(t)}\frac{dt}{t}\right].
\end{eqnarray}
One implication of the bootstrap formalism is that all breather form factors
can be deduced from the soliton-antisoliton form factors. In particular, the
form factor for the $k${\it th} breather ($k$ being odd) is given by:
\begin{equation}
\langle 0| j(0,0)|\theta\rangle_k = 2M_s\xi {\rm
cosh}(\theta){\prod_{l=1}^{k}{\rm tan}\frac{\pi l\xi}{2}\over \left({\rm
tan}\frac{\pi k\xi}{2}\right)^{1/2}}F(ik\xi-i\pi).
\label{breatherFF}
\end{equation}
Similarly, the two breathers form factor can be obtained from the four
soliton-antisoliton one. The procedure for its calculation has been
described by Smirnov. Following the procedure, we obtained the following
result (we assume that $k_1$ is odd and $k_2$ even):
\begin{widetext}
\begin{eqnarray}
 \langle 0| j(0,0) |\theta_1,\theta_2\rangle_{k_1,k_2} &=& 4\xi M_s
d_{k_1}d_{k_2}\zeta_{k_1,k_2}(\theta_{-}) 
\\ [3mm] \nonumber
& \times & \sum_{j=0}^{k_2} \Biggl[ (-1)^j {\prod_{l=1}^{k_1-1}{\rm
cosh}\frac{1}{2}\left( \theta_{-}-\frac{i\pi\xi}{2}(k_1+k_2-2j-2l)\right)
\over \prod_{l=0}^{k_1}{\rm sinh} \frac{1}{2}\left(
\theta_{-}+\frac{i\pi\xi}{2}(k_1-k_2+2j-2l)\right)}
 {\prod_{l=1}^{k_2-1}{\rm cos}\frac{\pi\xi}{2}(l-j) \over \prod_{l=0, \neq
j}^{k_2} {\rm sin} \frac{\pi\xi}{2}(l-j)}{\rm cosh}(\theta_{+})\Biggr] \; ,
\label{breather-breatherFF}
\end{eqnarray}
where
\begin{eqnarray}
 d_k = \left( {k{\rm cos}^2\left(\frac{\pi\xi k}{2}\right)\over 2\xi^3 {\rm
sin}\frac{\pi}{\xi}}\right)^{1/2}
 {\rm exp}\left\{ -\int_0^\infty\left[ {\rm sinh}^2((2-\xi k)t)\over {\rm
sinh}(2t)- {{\rm sinh}((1-\xi k)2t)\over 2}\right]{{\rm sinh}((1-\xi)t)\over
{\rm cosh}(t){\rm sinh}(\xi t)} {dt\over t} \right \}\quad ,
\end{eqnarray}
and
\begin{eqnarray}
\zeta_{k_1,k_2}(\theta) &=& c_{k_1,k_2} \frac{2\prod_{j=0}^{|k_1-k_2|}{\rm
sinh}\frac{1}{2}\left(\theta+\frac{i\pi\xi}{2}(|k_1-k_2|-2j)\right)}
{\prod_{l=1}^{k_1+k_2-1}{\rm cosh}\frac{1}{2}\left(\theta+\frac{i\pi\xi}{2}(k_1+k_2-2l)\right)}
\nonumber \\
& \times & {\rm exp}\left\{ -4\int_0^\infty \frac{ {\rm sinh}^2((1-i\theta/\pi)t){\rm cosh}((1-\xi k_1) t){\rm sinh}(\xi k_2t){\rm cosh}(\xi t)} 
{{\rm sinh}(\xi t){\rm
cosh}(t){\rm sinh}(2t)} \frac{dt}{t} \right\} \quad ,
\end{eqnarray}
\begin{eqnarray}
 c_{k_1,k_2} = 
{\rm exp}\left\{2\int_0^\infty {k_2{\rm cosh}(t){\rm sinh}(\xi t)-{\rm
cosh}((1-\xi k_1)t) {\rm sinh}(\xi k_2 t){\rm cosh}(\xi t)\over {\rm
sinh}(2t){\rm sinh}(\xi t){\rm cosh}(t)} \frac{dt}{t} \right\} \quad .
\end{eqnarray}
\end{widetext}
Using the obtained form factors, the different contributions to the
conductivity are shown in Figs~\ref{sol-solconduc} and \ref{b1-b2conduc}. Fig.~\ref{sol-solconduc}
shows the soliton-antisoliton contribution to the optical conductivity. This
contribution dominates at frequencies close to $2M_s$.

\end{document}